\documentclass[journal]{IEEEtran}
\usepackage{float}
\usepackage{amsmath,amsfonts}

\usepackage{array}
\usepackage{textcomp}
\usepackage{stfloats}
\usepackage{url}
\usepackage{verbatim}
\usepackage{graphicx}
\usepackage{cite}
\usepackage{multirow}
\usepackage{booktabs}
\usepackage{color}
\usepackage{graphicx}
\usepackage{amsmath}
\usepackage{amssymb}
\usepackage[section]{placeins}
\usepackage{float}
\usepackage{diagbox}
\usepackage{algpseudocode}
\usepackage{amsmath}
\usepackage{epstopdf}
\usepackage{amsmath}
\usepackage{amssymb}
\usepackage{multirow}

\usepackage{xcolor}
\usepackage{colortbl}
\usepackage{xcolor}
\usepackage{graphicx}%
\usepackage{multirow}%
\usepackage{amsmath,amssymb,amsfonts}%
\usepackage{amsthm}%
\usepackage{mathrsfs}%
\usepackage{xcolor}%
\usepackage{textcomp}%
\usepackage{manyfoot}%
\usepackage{booktabs}%
\usepackage{algorithm}%
\usepackage{algorithmicx}%
\usepackage{algpseudocode}%
\usepackage{listings}%
\usepackage{cite}
\usepackage{hyperref}
\usepackage{array}
\usepackage[caption=false,font=scriptsize,labelfont=sf,textfont=sf]{subfig}
\usepackage{stfloats}
\usepackage{url}
\usepackage{verbatim}
\usepackage{color}
\usepackage{cleveref}
\usepackage[section]{placeins}
\usepackage{float}
\usepackage{diagbox}
\usepackage{epstopdf}
\usepackage [utf8]{inputenc}
\usepackage{bbding}
\usepackage{booktabs}
\usepackage{makecell}
\usepackage{tabularx}
\usepackage{tikz}
\usepackage{algorithm}
\usepackage{ltxtable}
\usepackage{microtype}
\usepackage{wrapfig}
\usetikzlibrary{shapes.geometric, arrows}
\usepackage{setspace}

\usepackage{url}

\ifCLASSINFOpdf
\else
\fi
\begin{document}

%

\title{Multimodal Reasoning with LLM for Encrypted Traffic Interpretation: A Benchmark}
%
%
%
%

\author{
Longgang Zhang, Xiaowei Fu, Fuxiang Huang, and Lei Zhang,~\IEEEmembership{Senior Member,~IEEE}
\thanks{This work was partially supported by National Natural Science Fund of China under Grants 92570110 and 62271090, Chongqing Natural Science Fund under Grant CSTB2024NSCQ-JQX0038, and National Youth Talent Project. \textit{(Corresponding author: Lei Zhang)}}
\thanks{\IEEEcompsocthanksitem L. Zhang, X. Fu and L. Zhang are with the School of Microelectronics and Communication Engineering, Chongqing University, Chongqing 400044, China.
(E-mail: zlg502361@gmail.com, xwfu@cqu.edu.cn, leizhang@cqu.edu.cn,)

Fuxiang Huang is with the School of Data Science, Lingnan University, Hong Kong, China.
(E-mail: fxhuang1995@gmail.com)
}
\thanks{Manuscript received April 19, 2015; revised August 16, 2015.}}

\markboth{Journal of \LaTeX\ Class Files,~Vol.~14, No.~8, August~2015}%
{Shell \MakeLowercase{\textit{et al.}}: Bare Demo of IEEEtran.cls for Computer Society Journals}
%



\IEEEtitleabstractindextext{%
\begin{abstract}

Network traffic, as a key media format, is crucial for ensuring security and communications in modern internet infrastructure. While existing methods offer excellent performance, they face two key bottlenecks: (1) They fail to capture multidimensional semantics beyond unimodal sequence patterns. (2) Their ``black box" property, i.e., providing only category labels, lacks an auditable reasoning process. We identify a key factor that existing network traffic datasets are primarily designed for classification and inherently lack rich semantic annotations, failing to generate human-readable evidence report. To address data scarcity, this paper proposes a Byte-Grounded Traffic Description (BGTD) benchmark for the first time, combining raw bytes with structured expert annotations. BGTD provides necessary behavioral features and verifiable chains of evidence for multimodal reasoning towards explainable encrypted traffic interpretation. 
Built upon BGTD, this paper proposes an end-to-end traffic-language representation framework (mmTraffic), a multimodal reasoning architecture bridging physical traffic encoding and semantic interpretation. In order to alleviate 
modality interference and generative hallucinations, mmTraffic adopts a jointly-optimized perception-cognition architecture. By incorporating a perception-centered traffic encoder and a cognition-centered LLM generator, 
mmTraffic achieves refined traffic interpretation with guaranteed category prediction. 
Extensive experiments demonstrate that mmTraffic autonomously generates high-fidelity, human-readable, and evidence-grounded traffic interpretation reports, while maintaining highly competitive classification accuracy comparing to specialized unimodal model (e.g., NetMamba). The source code is available at \href{https://github.com/lgzhangzlg/Multimodal-Reasoning-with-LLM-for-Encrypted-Traffic-Interpretation-A-Benchmark}{Traffic-Reasoning-Project}.

\end{abstract}

\begin{IEEEkeywords}
Encrypted traffic classification, network traffic interpretation, large language model, multimodal learning.
\end{IEEEkeywords}}

\maketitle

\IEEEdisplaynontitleabstractindextext

%
\IEEEpeerreviewmaketitle

\section{Introduction}\label{sec1}
\IEEEPARstart{N}{etwork} traffic analysis is a core pillar for ensuring network security, implementing intrusion detection, and conducting traffic engineering. With the widespread deployment of Transport Layer Security (TLS 1.3), Quick UDP Connections (QUIC), and anonymous routing networks such as Tor~\cite{tor}, end-to-end encryption has made payload content extremely opaque. This evolution has rendered traditional Deep Packet Inspection (DPI) mechanisms, relying on plaintext signature matching, largely ineffective. Facing this challenge, encrypted traffic classification techniques have emerged.
These methods heavily relied on statistical features (e.g., packet size distribution and arrival time intervals, etc.) and machine learning techniques, but struggled to adapt to the highly-dimensional and dynamic adversarial nature of modern network traffic. 
In contrast, deep learning (DL) models achieved significant performance improvements by automatically extracting hierarchical representations from raw byte sequences. In recent years, inspired by the success of self-supervised pre-training in large models, traffic analysis models based on Transformers~\cite{transformer} and state space models (SSMs)~\cite{mamba} are emerged. For example, ET-BERT~\cite{ET_BERT} introduced a masked burst flow model, MPAF~\cite{MPAF} proposed a multi-phase attribute fingerprint, 
YaTC~\cite{YATC} proposed a multi-level flow representation (MFR) matrix, NetMamba~\cite{NetMamba} achieved ultra-fast inference using the linear-time complexity of the Mamba architecture, FlowletFormer~\cite{flowletformer} further optimized alignment capabilities by introducing behavior-semantic-aware Flowlet units, and WF-Transformer~\cite{WF-Transformer} further proposed a Transformer-based temporal feature extraction method. 

\begin{figure}[t]
  \centering
  \includegraphics[width=0.95\linewidth]{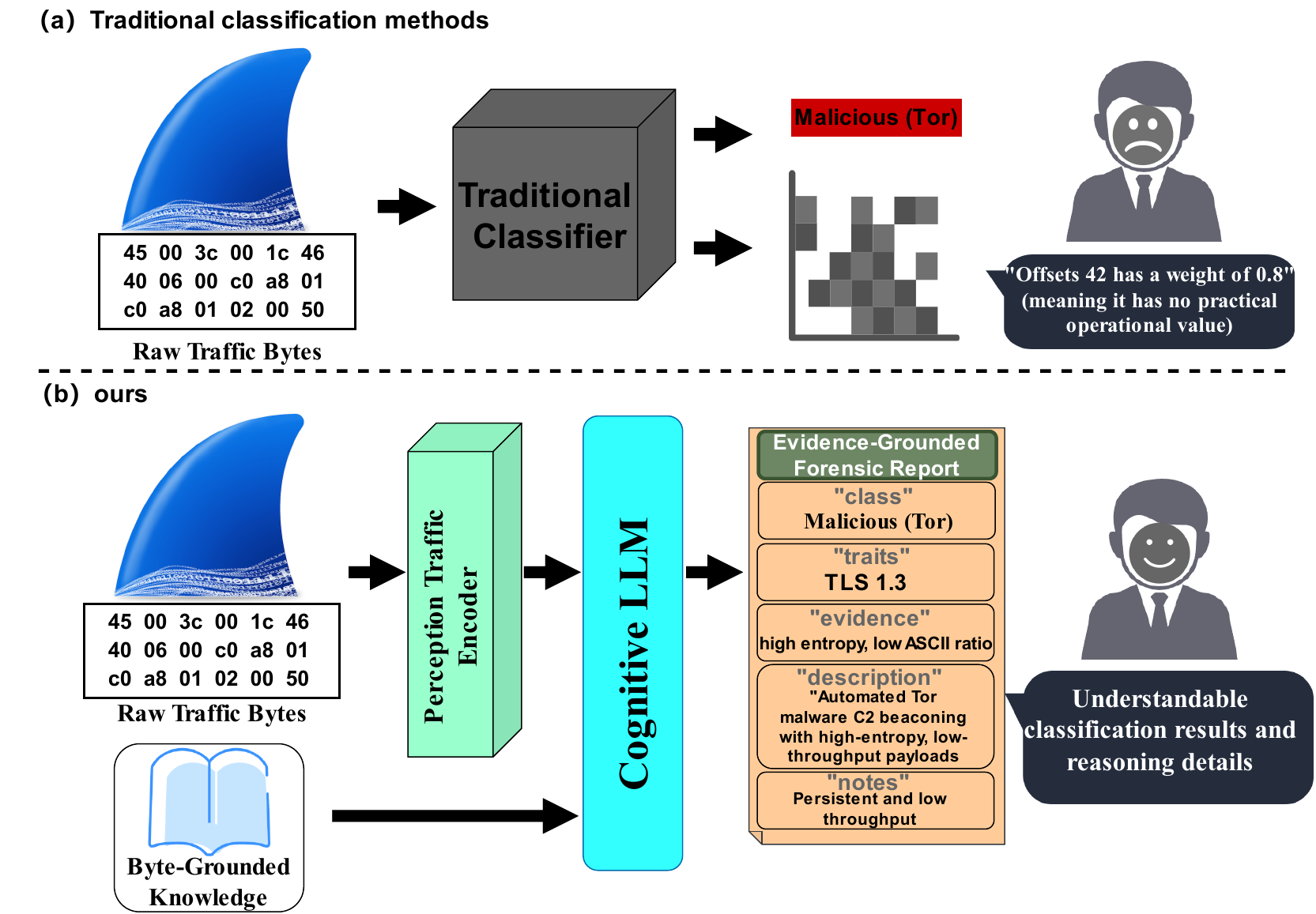}
  \caption{Comparison of traffic analysis paradigms. (a) Traditional classification methods that act as a ``black box", providing only a label and low-level feature weights that lack operational value. (b) Our proposed multimodal reasoning framework, composed of a Traffic Perception Encoder and a Cognitive LLM, instructed by Byte-Grounded Knowledge, generating an evidence-grounded report with human-understandable reasoning and executable insights.}
  \label{fig:introd}
\end{figure}

Despite the empirical success of deep representation learning models, contemporary cryptographic traffic analysis models remain constrained by two key bottlenecks: (1) \textit{Semantic Void in Unimodal Representations}. Existing models essentially perform nonlinear boundary partitioning in a high-dimensional space, directly mapping pure numerical hexadecimal byte sequences to classification labels. In complex enterprise environments, security analysts often encounter the ``statistical twin" phenomenon, i.e., benign traffic and malicious traffic employing obfuscation techniques exhibit almost identical statistical distributions. Relying solely on unimodal sequence patterns makes these models inadequate to capture the rich, multidimensional semantics required to distinguish such threats. (2) \textit{Black-box Property and Limitations of Traditional Explainable Artificial Intelligence (XAI)}. Purely statistical classifiers cannot provide human-readable, auditable, protocol-level forensic evidence to justify their decisions. While post-hoc interpretation techniques (e.g., SHAP~\cite{SHAP}, LIME~\cite{LIME} and Grad-CAM~\cite{GradCAM}) attempted to address this, they can only generate importance scores for features or attention heatmaps. For frontline Security Operations Center (SOC) analysts, knowing that ``the byte with offset 42 has high weight" is of no operational value unless the byte can be logically mapped to a specific protocol anomaly, such as a malformed handshake frame or an illegal cipher suite.

To overcome the aforementioned semantic limitations and black-box constraints, the deep model is expected to learn to map low-level physical bytes to high-level protocol semantics. \textit{However, existing network traffic datasets are primarily collected for the traditional classification task, providing only discrete category labels and inherently lacking the rich, multi-dimensional semantic annotations required, and thus unable to train generative interpretable models}. To bridge this fundamental gap, we innovatively construct a Byte-Grounded Traffic Description (BGTD) dataset. To the best of our knowledge, \textit{BGTD is the first benchmark that explicitly pairs raw network traffic bytes with structured, rich expert knowledge}. To ensure strong generalization capabilities, the dataset integrates six authoritative public repositories covering a broad ecosystem of applications. Beyond basic classification, BGTD provides fine-grained semantic annotations such as discriminative behavioral features, verifiable chains of evidence, and natural language descriptions. These elements are constructed through an automated expert knowledge generation process powered by \texttt{Claude Opus}. By linking numerical payloads with the high-level forensic information, BGTD provides the key foundational data required for multimodal reasoning.

Building upon this multimodal benchmark, this paper proposes an end-to-end, multi-modal traffic-language representation framework (\textbf{mmTraffic}) to overcome the inherent limitations of semantic void and black-box property in traditional traffic classifiers. Unlike traditional pipelines that strictly freeze the traffic encoder to prevent catastrophic forgetting and often lead to weak semantic alignment, mmTraffic advocates for a joint optimization for perception and cognition modules. By introducing an auxiliary classification head in perception and a semantic-priority guided generation mechanism in cognition, our framework explicitly constrains the continuous feature space and forces the large language model (LLM) to perform accurate classification before reasoning. This intrinsically empowers LLM to understand non-semantic traffic bytes and generate human-readable, evidence-grounded reports.

Fig.~\ref{fig:introd} describes the paradigm difference between mmTraffic and others. The main contributions are summarized as follows:

\begin{itemize}
    \item \textbf{A Byte-grounded traffic description benchmark (BGTD).} We construct the first benchmark to explicitly pair raw network traffic bytes with structured expert knowledge. By providing discriminative behavioral traits and verifiable chains of evidence, BGTD bridges the fundamental data-knowledge gap and enables multimodal reasoning towards interpretable encrypted traffic analysis.
    
    \item \textbf{A multi-modal traffic reasoning framework (mmTraffic).} We reformulate encrypted traffic analysis as a jointly optimized multimodal alignment pipeline. By unfreezing the traffic encoder and training it synergistically with the LLM, we achieve a deep semantic mapping from physical network bytes to human-readable concepts.
    
    \item \textbf{Auxiliary constraint and semantic-priority generation.} We introduce a classification head to enforce discriminative constraints on the traffic encoder. Furthermore, we design a semantic-priority generation loss that dynamically assigns higher weights to the categorical tokens, effectively mitigating LLM hallucinations in category prediction and ensuring the quality of generated reports. 
    
    \item \textbf{Superior performance of traffic interpretation with classification.} Extensive evaluations across six diverse traffic benchmarks demonstrate that \textbf{mmTraffic} achieves high-fidelity, auditable report generation, while maintaining exceptional classification accuracy. 
\end{itemize}

\section{Related Work}
\subsection{Self-supervised Methods for Encrypted Traffic Classification}
Large-scale self-supervised representation learning for network traffic is one of the most significant breakthroughs in cybersecurity in recent years. Early efforts primarily adapted paradigms from natural language processing and computer vision. For instance, ET-BERT~\cite{ET_BERT} pioneered the application of transformer architectures to traffic sequences via binary segmentation and masked burst flow modeling. Conversely, YaTC structured raw traffic as a multi-level flow representation (MFR) matrix, employing a dual-attention masked autoencoder to explicitly capture hierarchical packet interactions. To address computational bottlenecks and structural limitations, recent research has shifted towards efficiency and behavioral semantics. NetMamba~\cite{NetMamba} innovatively introduced state-space model (SSM)~\cite{mamba} via a stride-based representation, achieving faster inference suitable for high-speed networks. Meanwhile, FlowletFormer\cite{flowletformer} moved beyond fixed-length truncation by encoding explicit multi-layer protocol semantics based on coherent behavioral interaction units. Beyond masked modeling paradigms, contrastive learning has also been explored as a self-supervised pre-training strategy for encrypted traffic analysis. For instance, SmartDetector~\cite{ETCL} proposes a Semantic Attribute Matrix (SAM) representation and designs a traffic data augmentation method to improve robustness against obfuscation strategies such as dummy packet injection, pre-training the detection model via contrastive learning to learn deep representations from unlabeled traffic data.

Despite the diverse architectures and continuous breakthroughs~\cite{traffic_review} in accuracy, these models share a \textbf{fundamental limitation}: they are entirely constrained by the nature of unimodal black-box classifiers, as shown in Tables~\ref{tab:model_comparison}. While they excel at the classification task, they can only map numerical sequences to discrete labels, but fail to reasoning and generate interpretable reports with chains of evidence.

\begin{table*}[t]
\centering
\caption{Comparisons of different paradigms for network traffic analysis. MBM, SBP, MAE, MFM, and FPT represent Masked Byte Model, Segment Burst Prediction, Masked Autoencoder, Masked Flow Model, and Flow Prediction Task, respectively.}
\label{tab:model_comparison}
\footnotesize
\renewcommand{\arraystretch}{1.3}
\begin{tabularx}{\textwidth}{@{}
  >{\raggedright\arraybackslash}p{2.2cm}
  >{\raggedright\arraybackslash}p{3.2cm}
  >{\raggedright\arraybackslash}p{2.8cm}
  >{\raggedright\arraybackslash}p{2.2cm}
  >{\raggedright\arraybackslash}X
  @{}}
\toprule
\textbf{Model} &
\textbf{Traffic Representation} &
\textbf{Core Structure} &
\textbf{Pre-training} &
\textbf{Limitations} \\
\midrule
\textbf{ET-BERT}~\cite{ET_BERT}
  & 4-hex Bigram / Burst Segmentation
  & Transformer Encoder
  & MBM / SBP
  & Ignores protocol hierarchy; uses natural language subword tokenization \\
\midrule
\textbf{YaTC}~\cite{YATC}
  & Multi-level Flow Representation (MFR) Matrix
  & Dual-Attention Transformer
  & MAE (Matrix Masking)
  & Fixed matrix dimensions; truncates long-range session features \\
\midrule
\textbf{NetMamba}~\cite{NetMamba}
  & Stride-based Byte Sequence
  & Unidirectional Mamba (SSM)
  & Masked Stride Reconstruction
  & Purely numerical mapping; lacks interpretability \\
\midrule
\textbf{FlowletFormer}~\cite{flowletformer}
  & Flowlet Behavioral Unit / Field Tokenization
  & Transformer Encoder
  & MFM / FPT
  & Black-box classifier; unable to output forensic reasoning \\
\bottomrule
\end{tabularx}
\end{table*}

\subsection{LLMs for Network Security}
Early applications of Large Language Models (LLMs) in cybersecurity were primarily limited to plain text tasks, such as threat intelligence aggregation~\cite{securebert}, log parsing, and vulnerability description summarization. However, recent study begun to explore domain-specific LLMs capable of directly interpreting underlying telemetry data. TrafficLLM\cite{trafficllm} represents a significant attempt to bridge the modality gap. It employs a traffic-domain tokenizer to compress protocol fields by reducing token length to an approximately half. 
While TrafficLLM\cite{trafficllm} has demonstrated the feasibility of feeding continuous/discrete telemetry data into an LLM, this one-tower early fusion architecture suffers from an inherent structural vulnerability. Forcing an LLM to simultaneously process discrete natural language tokens and high-entropy, non-semantic numerical traffic tokens within the same attention layers frequently induces modality interference. Consequently, in high-risk intrusion detection, this architecture may neglect the authenticity of underlying physical bytes in order to maintain the fluency of the language, inevitably generating fictitious security alert logic. 
In contrast, the proposed mmTraffic explicitly mitigates this limitation by reformulating the architecture as an end-to-end multimodal framework, 
fundamentally bridges the modality gap, prevents generative hallucinations, and forces the LLM to ground its reasoning in authentic physical bytes.

\subsection{Multimodal Alignment and Cross-Modal Fusion}
The problem of bridging heterogeneous modalities is well-studied in the vision-language domain. Early approaches to cross-modal alignment include graph-based relational modeling~\cite{han2024tmm} and semantic-driven hashing for large-scale retrieval~\cite{lu2020tmm}, which established the importance of preserving semantic correspondences across modalities. CLIP~\cite{clip} demonstrated that contrastive alignment between image and text encoders produces powerful transferable representations. Subsequent works such as LLaVA~\cite{llava} and InstructBLIP~\cite{instructblip} extended this paradigm by using lightweight projection connectors to map frozen visual encoders into the token space of large language models, enabling instruction-following behavior over visual inputs. Flamingo~\cite{flamingo} further showed that cross-modal fusion via gated attention layers enables few-shot generalization across diverse vision-language tasks. mmTraffic draws the following insight: \textit{rather than relying on disparate training stages with a frozen perception module, we align an active traffic encoder with a language model through a lightweight MLP connector}, empowering the LLM to perform encrypted traffic interpretation with rigorous multimodal reasoning.

\subsection{Explainability in Traffic Analysis}
Despite the strong empirical performance of deep traffic classifiers, their black-box nature has motivated a growing body of work on explainable AI (XAI)~\cite{xAI}. Post-hoc techniques such as SHAP~\cite{SHAP}, LIME~\cite{LIME}, and Grad-CAM~\cite{GradCAM} provide feature-level attribution scores, but cannot produce protocol-level forensic evidence for security analysts. While attention-based mechanisms have been extended to model inter-modal interactions~\cite{liu2023tmm} and structured multimodal representations~\cite{guo2020tmm}, these approaches remain confined to feature-level enhancement without producing human-readable explanations. DISTILLER~\cite{DISTILLER} proposed a multimodal multitask framework that jointly learns traffic representations and human-readable labels, but still lacks free-form natural language generation. mmTraffic addresses this gap by leveraging large language models to produce structured, evidence-grounded forensic reports, moving beyond importance scores toward auditable reasoning chains.

\section{BGTD Benchmark for Traffic Reasoning}
\subsection{Overview}

A benchmark that explicitly links raw traffic analysis data with expert-level semantic reasoning is a prerequisite for training a multimodal traffic reasoning framework, but is still unexplored. Therefore, we develop a Byte-Grounded Traffic Description (BGTD) benchmark, which, to the best of our knowledge, bridges the data scarcity and persistent data-knowledge gap in encrypted traffic interpretation for the first time.
To ensure the diversity of data distribution and scenarios, the BGTD dataset integrates six authoritative public traffic repositories, covering different network behaviors, application ecosystems, and encryption protocols. Specifically, BGTD deeply integrates cross-platform mobile application traffic (i.e., CrossPlatform-Android~\cite{crossplatform} and CrossPlatform-iOS\cite{crossplatform}), cutting-edge TLS 1.3 encrypted web communication (i.e., CSTNet-TLS1.3)~\cite{ET_BERT}, complex encrypted VPN tunnels and anonymous routing networks (i.e., ISCXVPN2016~\cite{ISCXVPN2016} and ISCX-Tor-2016)~\cite{ISCXTor2016}, and hybrid malware traffics containing multiple attack families (i.e., USTC-TFC-2016~\cite{USTC-TFC2016}). The pipeline for the BGTD dataset is shown in Figure~\ref{fig:dataset_generate}.

\subsection{Session Extraction and Class Balancing}

\begin{figure*}[t]
  \centering
  \includegraphics[width=0.95\textwidth]{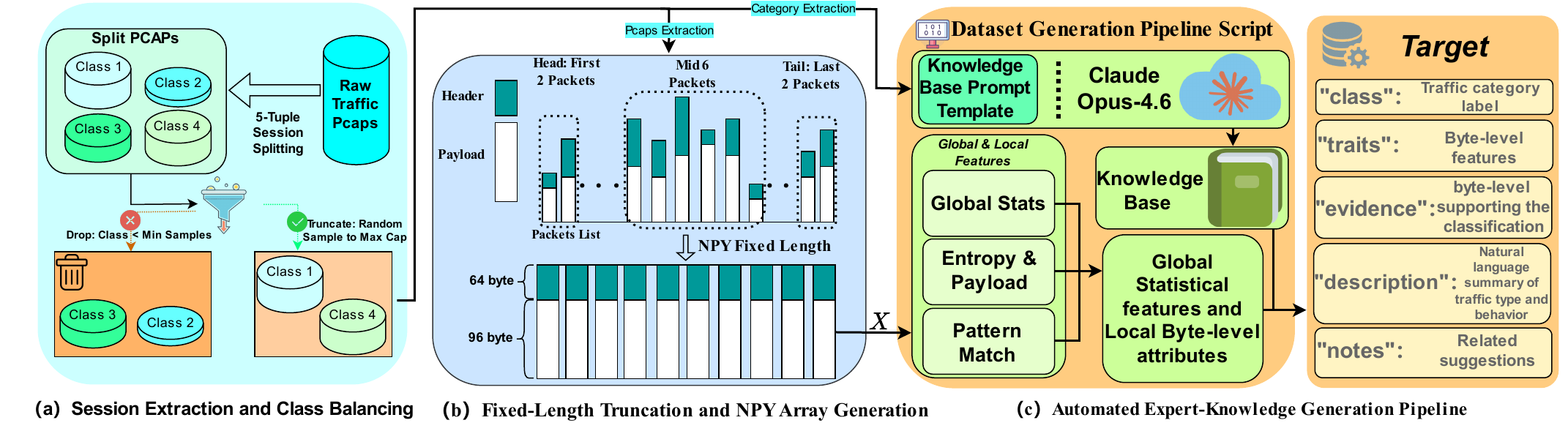}
  \caption{Pipeline of developing  BGTD dataset: (a)~session extraction and class balancing from raw PCAP files, (b)~fixed-length $10\times160$ NPY array generation via priority-based packet sampling, and (c)~LLM-assisted ground-truth synthesis using \texttt{Claude Opus-4.6} prompted as a senior network security expert.}
  \label{fig:dataset_generate}
\end{figure*}

As shown in Figure~\ref{fig:dataset_generate} (a), the raw PCAP files from various datasets undergo a multi-stage preprocessing. First, each PCAP file is partitioned into the standard five-tuple format (i.e., source IP address, destination IP address, source port, destination port, protocol). To mitigate the impact of the long-tail distribution, the original dataset is filtered by category, with lower and upper sample thresholds applied. Categories below the lower threshold are removed, while categories above the upper threshold are sampled according to the threshold. Specific processing methods for each dataset are provided in Sec.~\ref{data pre}. The statistics of the BGTD dataset are shown in Figure~\ref{fig:dataset_stats}.

\subsection{Fixed-Length Truncation and NPY Array Generation}

Each segmented flow from above step is treated as an independent sample. To extract informative byte-level features from the original traffic data, we implement a heuristic priority-based sampling algorithm, aiming to transform variable-length network flows into fixed-dimensional tensor features. This algorithm does not employ simple sequential truncation or random sampling, but rather comprehensively considers the temporal structure and payload information of the flow. Its specific execution logic is as follows:

\begin{figure}[t]
  \centering
  \includegraphics[width=\linewidth]{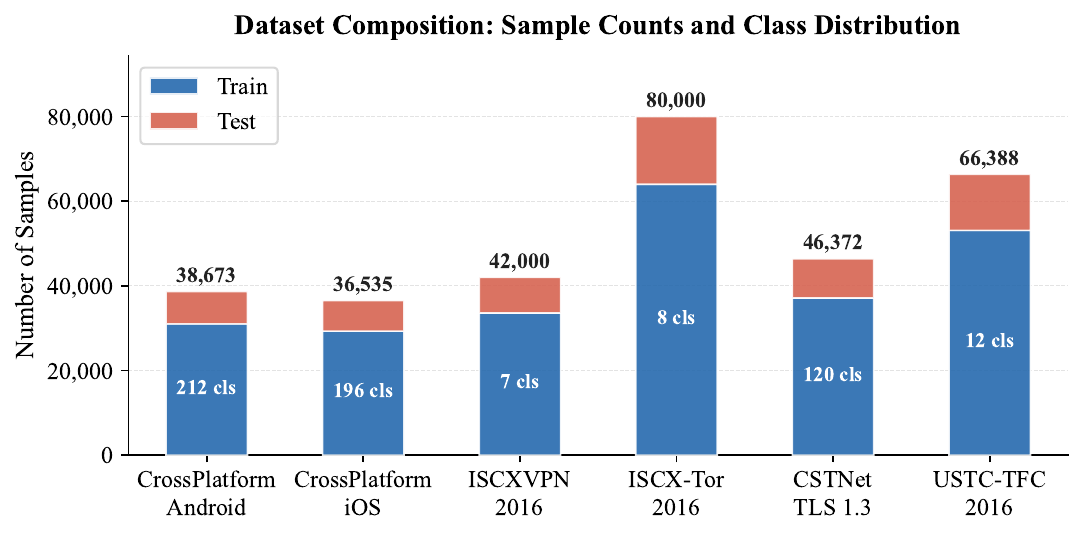}
  \caption{Statistical overview of the BGTD dataset.}
  \label{fig:dataset_stats}
\end{figure}

\begin{itemize}
	\item \textbf{Temporal Keyframe Preservation:} The algorithm forcibly preserves the first two packets and the last two packets in the flow sequence. Preserving the header packets is to capture key metadata such as protocol handshakes and control negotiations. Preserving the tail packets helps record the state characteristics. 
    \item \textbf{Payload Information Filtering:} For packets in the middle of the stream, the algorithm sorts them in descending order based on the transport layer ($L4$) payload length and prioritizes packets with larger payloads to fill the preset sampling number $K$ ($K=10$ in this paper). This strategy is based on the assumption that packets with larger payloads typically carry richer application-layer protocol fingerprints.
    \item \textbf{Dimension Alignment and Consistency Guarantee:} To ensure every sample has exactly $K$ packets, the algorithm applies two strategies depending on the flow length $n$. If $n \geq K$, the first two and last two packets are retained as structural anchors, and the remaining $K{- }4$ slots are filled by selecting packets from the middle region in descending order of payload size, prioritizing information-rich packets. If the quota is still unmet, equidistant indices are computed via $\mathtt{linspace}(0, n{-}1, \text{deficit})$ and the corresponding packets are appended. If $n < K$, all packets are kept and the sequence is extended by cyclic repetition, i.e., position $j$ maps to packet $j \bmod n$, until exactly $K$ packets are obtained.
\end{itemize}

Simultaneously, to ensure the consistency of the input dimensions for subsequent processing and focus on the critical protocol negotiation phase, each packet obtained by the packet sampling algorithm is truncated or padded to a fixed length of 160 bytes. As shown in Figure~\ref{fig:dataset_generate} (b), the 160 bytes consists of a 64-byte header area (i.e., $L3/L4$ header used to capture protocol metadata) and a 96-byte payload area. Ultimately, each network stream is transformed into a fixed-dimensional tensor $X \in \mathbb{R}^{10 \times 160}$ containing a custom protocol ID byte, a 63-byte processed header, and a 96-byte payload, which is then flattened into a 1600-dimensional continuous byte matrix. Furthermore, to protect privacy and prevent the model from overfitting specific network identifiers, the system forcibly masks the source and destination IP addresses at the network layer and uses a bucketing mechanism to map transport layer ports into three categories: privileged ports, registered ports, and dynamically private ports. This de-identification strategy forces the model to focus on protocol semantics and payload sequence patterns, thereby improving the model's generalization ability. Ultimately, for ease of development, the pre-processed traffic data is stored as NPY array.

\subsection{Automated Expert-Knowledge Generation Pipeline} 

To construct relationships between traffic data and reliable analysis reports, we resort to large-scale language models to conduct a structured process from low-level data to high-level semantics~\cite{chatgpt}, as shown in Figure~\ref{fig:dataset_generate} (c).

\textbf{Global and Local Feature Extraction.} Firstly, to capture interpretable reasoning evidence, we deploy a Dataset Generation Pipeline Script to extract global statistical features and local byte-level attributes from traffic analysis data $X$:
\begin{itemize}
	\item \textit{Global Statistical Features:} This data includes the duration of traffic, average packet size, throughput (Bps), and the proportion of dominant protocols. These features are transformed into semantic descriptions by script. For example, an average packet length exceeding 800 bytes is mapped to ``\texttt{large-volume data transmission characteristics}".
	\item \textit{Encryption and Payload Distribution Evaluation:} For encrypted traffic, such as TLS 1.3 environments, the Shannon entropy of non-zero payload regions and the proportion of printable ASCII characters are calculated. Based on the statistical distribution of the entire dataset, these continuous indicators are discretized into three levels: low, mid, and high, at the 33rd and 66th percentiles. For example, ``low ASCII rate" combined with ``high Shannon entropy" will serve as a strong signal of encrypted or compressed data characteristics.
	\item \textit{Deterministic Pattern Matching:} Detect if the payload contains obvious plaintext HTTP methods (e.g., GET, POST, etc.) or TLS record layer header features (e.g., 0x14-0x17 with version number 0x03) via feature matching, providing hard logic support for classification.
\end{itemize}

\textbf{Expert Knowledge Base Construction.} Secondly, to address the lack of rich semantic descriptions in traditional traffic datasets, this study introduces a large language model (\texttt{Claude Opus-4.6}) to help construct a structured domain expert knowledge base. For each traffic category in the dataset, according to preset Knowledge Base Prompt Template, LLM automatically generates an expert description containing three dimensions: \textbf{(1)} a protocol hint that concisely defines the application or protocol to which the traffic belongs; \textbf{(2)} behavioral characteristics describing 3 to 5 typical patterns of the traffic at the network level; \textbf{(3)} a security context that provides supplementary explanations from a network security and traffic monitoring perspective, identifying the key distinguishable features 
among easily confused categories. This knowledge base provides powerful domain knowledge for subsequently constructing rich, multi-perspective training texts.

\textbf{Multi-field Semantic ``Label'' Generation.} Thirdly, upon the above features and expert knowledge base, the dataset generation pipeline conducts a structured process to generate fine-grained semantic labels (i.e., ``Target'' in Figure~\ref{fig:dataset_generate} (c)), comprising 5 structured fields.
The \textbf{class} field provides the ground-truth traffic category label, directly derived from the directory structure of the original dataset after session splitting and class balancing.
The \textbf{traits} field encodes five deterministic byte-level attributes extracted from the NPY array: a boolean indicating the presence of TLS record header patterns, a boolean indicating the presence of plaintext HTTP tokens, and three discretized bucket indicators for ASCII ratio, Shannon entropy, and zero-padding ratio, each categorized as \textit{low, mid, high} based on the 33rd and 66th percentiles of the full data distribution.
The \textbf{evidence} field contains 2 to 4 natural language statements constructed by combining the above byte-level traits and global features. Each statement describes a concrete, verifiable observation grounded in the raw byte data (e.g., \textit{``High Shannon entropy in non-zero payload regions indicates that the data is highly likely to have been encrypted or compressed''}).
The \textbf{description} field provides a 2 to 3 sentence behavioral summary that integrates byte-level observations with the expert knowledge base, depicting the protocol attribution, application-layer characteristics, and typical communication behavior of the traffic.
The \textbf{notes} field supplies a single security-relevant sentence drawn from the knowledge base's security context, highlighting potential misuse risks, recommended monitoring strategies, or distinguished indicators for anomaly detection.

Ultimately, all five fields are serialized together as a structured JSON object and stored in JSONL format, forming the complete training target for the proposed mmTrafficframework.

\section{The Proposed mmTraffic}
\subsection{Overview of the Framework}

\begin{figure*}[t]
  \centering
  \includegraphics[width=0.95\linewidth]{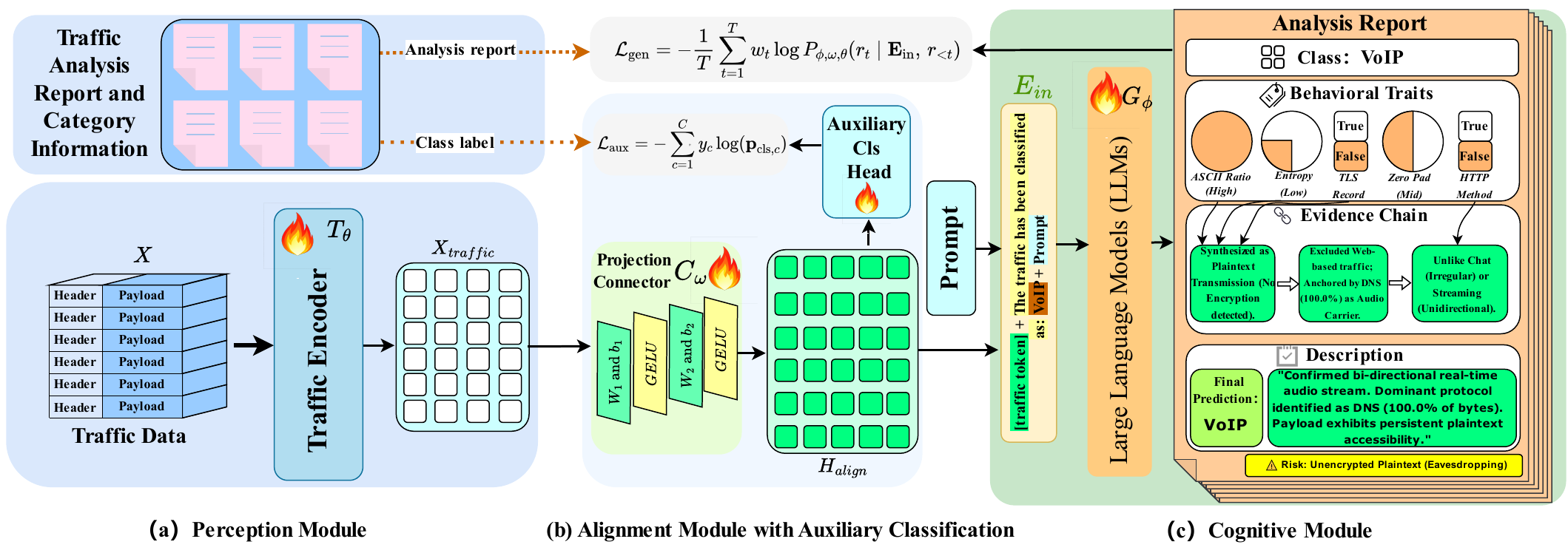}
  \caption{Overview of the mmTraffic framework. (a)~The frozen traffic encoder ${T}_{\theta}$ extracts high-dimensional features from raw traffic data. (b)~The linear connector ${C}_{\omega}$ projects traffic features into the LLM token space, with the CGHF mechanism injecting a class-aware anchor token into the input sequence. (c)~The LLM ${G}_{\phi}$ autoregressively generates a structured forensic report containing behavioral traits, evidence chain, and diagnostic description.}
  \label{fig:llava_model}
\end{figure*}

The pipeline of \textbf{mmTraffic} is illustrated in Figure~\ref{fig:llava_model}. It comprises 
three highly-collaborative and jointly-optimized modules:  \textbf{Perception} module, \textbf{Alignment} module, and \textbf{Cognition} module. First, the Perception module acts as the foundational feature extractor. Unlike previous paradigms that freeze the traffic encoder, our encoder actively participates in the multimodal training phase, updating its parameters to learn language-aligned representations directly from raw traffic bytes.
Second, the Alignment module bridges the dedicated traffic latent space and the natural language lexical space. To force this projected space to capture highly discriminative semantics autonomously, we introduce an auxiliary classification head with a dedicated constraint loss. This ensures the continuous features possess clear, linear-separable categorical boundaries before entering the language model. Finally, the Cognition module leverages the aligned multimodal embeddings to perform autoregressive reasoning. To ensure the generated traffic analysis report remains logically rigorous, we propose a Semantic-Priority Guided Generation mechanism. This mechanism dynamically assigns higher optimization weights to the categorical tokens generated at the beginning of the sequence, compelling the large language model (LLM) to perform accurate classification before reasoning about verifiable chains of evidence. By transitioning to an end-to-end joint optimization strategy, \textbf{mmTraffic} empowers the LLM to intrinsically understand and classify non-semantic traffic sequences, successfully achieving accurate classification and evidence-grounded interpretation within a unified framework.

\subsection{Perception Module}

This module receives the raw byte sequence $X$ preprocessed according to the BGTD protocol and performs a high-dimensional non-linear mapping via the traffic encoder $T_{\theta}$.

High-dimensional Continuous Feature Embedding: The encoder $T_{\theta}$ processes $X$ to automatically extract complex spatial dependencies and structural patterns of protocol fields, generating a dense feature tensor:
\begin{equation}
    \mathbf{T}_{traffic} = {T}_{\theta}(\mathbf{X})
\end{equation}
where $T_{traffic} \in \mathbb{R}^{L \times d_{traffic}}$, $L$ denotes the sequence length, and $d_{traffic}$ represents the feature dimension of the traffic encoder.

End-to-End Optimization: In contrast to previous multi-stage paradigms where the perception module is trained independently and then strictly frozen to prevent catastrophic forgetting, our framework treats $T_{\theta}$ as an active component within a unified multimodal architecture. The parameters of the traffic encoder are unfrozen and updated during the joint training phase. By receiving gradient feedback backpropagated from both the downstream auxiliary classification head $A_{\kappa}$ and the cognitive language model, the encoder is explicitly guided to map non-semantic raw bytes into a representation space that is naturally aligned with language-based reasoning.

\subsection{Alignment Module with Auxiliary Classification}

To achieve implicit alignment between the dedicated traffic latent space and the natural language lexical space, \textbf{mmTraffic} deploys a lightweight projection connector $\mathrm{C}_{\omega}$. For computational efficiency, we employ a two-layer Multi-Layer Perceptron (MLP) equipped with a Gaussian Error Linear Unit (GELU) activation. Given the continuous traffic embedding $\mathbf{T}_{\mathrm{traffic}}$ from the Perception module, the non-linear transformation of $\mathrm{C}_{\omega}$ is formally defined as:
\begin{equation}
\mathbf{H}_{\mathrm{align}} = \mathrm{C}_{\omega}(\mathbf{T}_{\mathrm{traffic}}) = \mathbf{W}_{2} \sigma(\mathbf{W}_{1} \mathbf{T}_{\mathrm{traffic}} + \mathbf{b}_{1}) + \mathbf{b}_{2}
\end{equation}
where $\sigma(\cdot)$ denotes the GELU activation function, and $\omega = \{\mathbf{W}_{1}, \mathbf{b}_{1}, \mathbf{W}_{2}, \mathbf{b}_{2}\}$ represents the learnable weight matrices and bias vectors of the connector. The resulting $\mathbf{H}_{\mathrm{align}}$ bridges the dimensional gap, projecting the traffic features into the LLM's lexical space.

\textbf{Auxiliary Classification.} During alignment, simply mapping the dimensions is insufficient to guarantee that the projected tokens carry explicit categorical semantics. To force the continuous feature space to capture highly discriminative information, we introduce an Auxiliary Classification Head, denoted as $\mathrm{A}_{\kappa}$, atop the projection connector.
Specifically, we first apply Global Average Pooling (GAP) across the sequence dimension $L$ of the aligned features $\mathbf{H}_{\mathrm{align}}$ to obtain a condensed sequence-level semantic representation $\mathbf{H}_{\mathrm{pool}} \in \mathbb{R}^{d_{\mathrm{h}}}$:
\begin{equation}
\mathbf{H}_{\mathrm{pool}} = \mathrm{GAP}(\mathbf{H}_{\mathrm{align}}) = \frac{1}{L} \sum_{i=1}^{L} \mathbf{H}_{\mathrm{align}}^{(i)}
\end{equation}
where $\mathbf{H}_{\mathrm{align}}^{(i)}$ represents the $i$-th token embedding in the sequence, and $d_{\mathrm{h}}$ is the hidden dimension of the LLM. This pooled representation is then processed by the auxiliary classification head $\mathrm{A}_{\kappa}$ to predict the discrete probability distribution $\mathbf{p}_{\mathrm{cls}} \in \mathbb{R}^{C}$ over $C$ predefined traffic classes:
\begin{equation}
\mathbf{p}_{\mathrm{cls}} = \mathrm{A}_{\kappa}(\mathbf{H}_{\mathrm{pool}}) = \mathrm{Softmax}(\mathbf{W}_{\mathrm{cls}} \mathbf{H}_{\mathrm{pool}} + \mathbf{b}_{\mathrm{cls}})
\end{equation}
where $\kappa = \{\mathbf{W}_{\mathrm{cls}}, \mathbf{b}_{\mathrm{cls}}\}$ are the learnable parameters of $\mathrm{A}_{\kappa}$, mapping the hidden dimension $d_{\mathrm{h}}$ to the category space $C$. 

Then we compute the auxiliary classification loss using the standard cross-entropy objective:
\begin{equation}
\mathcal{L}_{\mathrm{aux}} = -\sum_{c=1}^{C} y_{c} \log(\mathbf{p}_{\mathrm{cls}, c})
\end{equation}
where $y_{c} \in \{0, 1\}$ is the binary indicator of the ground-truth class label, and $\mathbf{p}_{\mathrm{cls}, c}$ is the predicted probability for class $c$. During the joint training phase, $\mathcal{L}_{\mathrm{aux}}$ actively propagates gradient back through $\mathrm{C}_{\omega}$ and $\mathrm{T}_{\theta}$, strictly anchoring the representation space to the identity of a traffic sample.

\subsection{Cognition Module}

With the auxiliary classification constraint enforcing category-aware features in the alignment stage, the Cognition module calls the large language model $\mathrm{G}_{\phi}$ to perform autoregressive inference directly from the aligned traffic features, without relying on hard-coded discrete labels. The full input sequence $\mathbf{E}_{\mathrm{in}}$ of $\mathrm{G}_{\phi}$ is constructed by concatenating the aligned traffic tokens $\mathbf{H}_{\mathrm{align}}$ and the task instruction prompt $\mathbf{P}$:
\begin{equation}
    \mathbf{E}_{\mathrm{in}} = [\mathbf{H}_{\mathrm{align}};\, \mathbf{P}]
\end{equation}
The LLM then autoregressively generates the diagnostic report $\mathbf{R}_{\mathrm{pred}}$ conditioned on this sequence:
\begin{equation}
    \mathbf{R}_{\mathrm{pred}} = \mathrm{G}_{\phi}(\mathbf{E}_{\mathrm{in}})
\end{equation}

In the multi-modal report generation task, the correct identification of the traffic category acts as the foundational premise for all subsequent behavioral descriptions and evidence chains. Standard negative log-likelihood (NLL) loss treats all generated tokens equally, which may lead to the LLM generating fluent but factually incorrect hallucinated reports if the core category is misidentified. To address this, we propose a Semantic-Priority Guided Generation Loss. We assign an amplification weight to the prefix tokens of the target sequence (which correspond to the primary categorical decision in the JSON structure) to explicitly force the LLM to prioritize classification accuracy during the generative process. The weighted generation loss over the expert evidence chains $\mathbf{R} = \{r_{1}, r_{2}, \dots, r_{T}\}$ is formulated as:
\begin{equation}
    \mathcal{L}_{\mathrm{gen}} = -\frac{1}{T}\sum_{t=1}^{T} w_{t} \log P_{\phi,\omega,\theta} (r_{t} \mid \mathbf{E}_{\mathrm{in}},\, r_{<t})
\end{equation}
where $r_{<t} = \{r_{1}, \dots, r_{t-1}\}$ denotes all ground-truth tokens preceding position $t$, and $w_{t}$ is the dynamic positional weight defined as:
\begin{equation}
    w_{t} = \begin{cases} 1 + \gamma, & \text{if } t \le M \\ 1, & \text{otherwise} \end{cases}
\end{equation}
where $M$ defines the boundary of the critical categorical tokens at the beginning of the sequence, and $\gamma$ is the boost weight factor applied to strictly penalize misclassifications in the generated text.


\begin{algorithm}[t]
\caption{The Training Pipeline of mmTraffic}
\label{alg:etlr}
\begin{algorithmic}[1]

\Require Traffic tensor $\mathbf{X} \in \mathbb{R}^{10 \times 160}$, ground-truth label $y$, chain of evidence annotation $\mathbf{R} = \{r_{1}, \ldots, r_{T}\}$ of BGTD, unfrozen encoder $\mathrm{T}_{\theta}$, connector $\mathrm{C}_{\omega}$, auxiliary classification head $\mathrm{A}_{\kappa}$, LLM $\mathrm{G}_{\phi}$, task prompt $\mathbf{P}$, auxiliary loss weight $\lambda$, semantic boost weight $\gamma$, and threshold $M$.
\Ensure Predicted traffic class $\hat{y}$ and forensic report $\mathbf{R}_{\mathrm{pred}}$.

\Statex \textbf{// Step 1: Perception Module}
\State $\mathbf{T}_{\mathrm{traffic}} \leftarrow \mathrm{T}_{\theta}(\mathbf{X})$ \Comment{Extract embeddings via unfrozen encoder}

\Statex \textbf{// Step 2: Alignment Module \& Auxiliary Constraint}
\State $\mathbf{H}_{\mathrm{align}} \leftarrow \mathrm{C}_{\omega}(\mathbf{T}_{\mathrm{traffic}})$ \Comment{Project to LLM lexical space}
\State $\mathbf{H}_{\mathrm{pool}} \leftarrow \mathrm{GAP}(\mathbf{H}_{\mathrm{align}})$ \Comment{Sequence-level global average pooling}
\State $\mathbf{p}_{\mathrm{cls}} \leftarrow \mathrm{A}_{\kappa}(\mathbf{H}_{\mathrm{pool}})$ \Comment{Predict auxiliary class distribution}
\State $\hat{y} \leftarrow \arg\max(\mathbf{p}_{\mathrm{cls}})$ \Comment{Traffic classification prediction}

\Statex \textbf{// Step 3: Multimodal Construction}
\State $\mathbf{E}_{\mathrm{in}} \leftarrow [\mathbf{H}_{\mathrm{align}};\; \mathbf{P}]$ \Comment{Directly concatenate features and prompt}

\Statex \textbf{// Step 4: Cognition Module \& Joint Optimization}
\If{training}
  \State $\mathcal{L}_{\mathrm{aux}} \leftarrow -\log(\mathbf{p}_{\mathrm{cls}, y})$ \Comment{Compute auxiliary cross-entropy loss}
  \State Compute dynamic weights $w_{t}$: $1+\gamma$ if $t \le M$, else $1$
  \State $\mathcal{L}_{\mathrm{gen}} \leftarrow -\frac{1}{T}\sum_{t=1}^{T} w_{t} \log P_{\phi,\omega,\theta}(r_{t} \mid \mathbf{E}_{\mathrm{in}}, r_{<t})$ \Comment{Semantic-priority guided loss}
  \State $\mathcal{L}_{\mathrm{total}} \leftarrow \mathcal{L}_{\mathrm{gen}} + \lambda \mathcal{L}_{\mathrm{aux}}$ \Comment{End-to-end joint objective}
  \State Update $\{\theta, \omega, \kappa, \phi\}$ via AdamW \Comment{Jointly optimize all modules}
\Else
  \State $\mathbf{R}_{\mathrm{pred}} \leftarrow \mathrm{G}_{\phi}(\mathbf{E}_{\mathrm{in}})$ \Comment{Autoregressive report generation}
  \State Parse $\mathbf{R}_{\mathrm{pred}}$ into JSON format: $\{class, traits, evidence, description, notes\}$
\EndIf
  \State \Return $\hat{y},\; \mathbf{R}_{\mathrm{pred}}$
\end{algorithmic}
\end{algorithm}

Unlike previous paradigms that freeze the traffic encoder, our mmTraffic framework enables end-to-end multimodal alignment. The overall objective function integrates both the token-level generative comprehension and the sequence-level auxiliary classification constraint:
\begin{equation}
    \mathcal{L}_{\mathrm{total}} = \mathcal{L}_{\mathrm{gen}} + \lambda \mathcal{L}_{\mathrm{aux}}
\end{equation}
where $\lambda$ is a hyperparameter balancing the auxiliary classification task. During training, the parameters of the traffic encoder ($\mathrm{T}_{\theta}$), projection connector ($\mathrm{C}_{\omega}$), auxiliary classification head ($\mathrm{A}_{\kappa}$), and language model ($\mathrm{G}_{\phi}$) are jointly optimized. 

In summary, the proposed mmTraffic constructs a closed-loop analysis system from the underlying bit stream to the high-level forensic report through a perception module (feature extraction), an alignment module (cross-space mapping with auxiliary constraint), and a cognition module (logical reconstruction with semantic-priority guidance). This joint optimization empowers the LLM to intrinsically understand non-semantic traffic sequences, ensuring the reliability of the analysis report. To clearly demonstrate the logical pipeline of the above procedure, we summarize \textbf{mmTraffic} in Algorithm~\ref{alg:etlr}.

\section{Experiments}
\subsection{Experimental Settings}
\subsubsection{Data Preparation}
\label{data pre}

\begin{table}[t]
\centering
\caption{Statistics of the six benchmark datasets after preprocessing. 
}
\label{tab:datasets}
\resizebox{\linewidth}{!}{%
\begin{tabular}{l c c c c c}
\toprule
\textbf{Dataset} & \textbf{Classes} & $N_{\min}$ & $N_{\max}$ & \textbf{Train} & \textbf{Test} \\
\midrule
CrossPlatform-Android & 212 & 50    & 2,000  & 31,029 & 7,644  \\
CrossPlatform-iOS     & 196 & 50    & 3,000  & 29,302 & 7,233  \\
ISCXVPN2016           & 7   & 200   & 6,000  & 33,600 & 8,400  \\
ISCX-Tor-2016         & 8   & 3,000 & 10,000 & 64,000 & 16,000 \\
CSTNet-TLS1.3         & 120 & 0     & 6,000  & 37,148 & 9,224  \\
USTC-TFC-2016         & 12  & 3,000 & 6,000  & 53,112 & 13,276 \\
\bottomrule
\end{tabular}%
}
\end{table}

We evaluate the proposed framework on six publicly available network traffic datasets, covering a wide range of traffic types, application ecosystems, and encryption protocols. To reduce the impact of class imbalance, we apply a category filtering strategy to each dataset: classes with fewer than $N_{\min}$ samples are removed, and classes exceeding $N_{\max}$ samples are randomly downsampled to $N_{\max}$. All datasets are split into training and test sets at an 8:2 ratio. Statistics of the six datasets after the above preprocessing are shown in Tables~\ref{tab:datasets}. Details of each dataset are as follows:

\begin{itemize}
    \item \mbox{\textbf{CrossPlatform-Android and CrossPlatform-iOS}}~\cite{crossplatform}: These two datasets contain mobile application traffic collected from Android and iOS devices across multiple countries and network environments, covering 212 and 196 application categories respectively. We set $N_{\min} = 50$, $N_{\max} = 2{,}000$ for Android and $N_{\max} = 3{,}000$ for iOS, resulting in 38,673 and 36,535 samples after filtering.
    
    \item \textbf{ISCXVPN2016}~\cite{ISCXVPN2016}: This dataset contains application traffic tunneled through VPN connections, covering 7 categories: \texttt{Browsing, Chat, Email, FTP, P2P, Streaming}, and \texttt{VoIP}. VPN encapsulation adds an additional encryption layer that obscures application-layer signatures. We set $N_{\min} = 200$ and $N_{\max} = 6{,}000$, yielding 42,000 samples.

    \item \textbf{ISCX-Tor-2016}~\cite{ISCXTor2016}: This dataset contains traffic routed through the Tor anonymous network, covering 8 traffic categories. Tor's multi-hop encryption substantially reduces the discriminability of byte-level features, making it one of the most challenging benchmarks in encrypted traffic analysis. We set $N_{\min} = 3{,}000$ and $N_{\max} = 10{,}000$, resulting in 80,000 samples.

    \item \textbf{CSTNet-TLS1.3}~\cite{ET_BERT}: This dataset contains web traffic encrypted exclusively with TLS 1.3, covering 120 website categories. Since TLS 1.3 removes observable handshake metadata, certificate-based identification is not applicable. The class distribution is relatively balanced, so $N_{\min} = 0$ and $N_{\max}=6,000$, yielding 46,372 samples.

    \item \textbf{USTC-TFC-2016}~\cite{USTC-TFC2016}: This dataset contains both benign application traffic and traffic from 12 malware families, making it the only dataset in our benchmark that includes adversarial network behavior. We set $N_{\min} = 3{,}000$ and $N_{\max} = 6{,}000$, resulting in 66,388 samples.
\end{itemize}

\begin{table*}[t]
\centering
\caption{Results on ISCX-Tor-2016, ISCXVPN2016, and CSTNet-TLS1.3. \textbf{NetMamba} is the unimodal classifier (``--'' means no generation); \textbf{Zero-shot LLM} feeds features directly to LLM without tuning; \textbf{Vanilla} freezes the encoder without auxiliary constraints. 
\textbf{Bold} indicates the best per column within each dataset group.}
\label{tab:main_results_1}
\resizebox{\linewidth}{!}{%
\begin{tabular}{l l c c cc cc}
\toprule
\multirow{2}{*}{\textbf{Dataset}} &
\multirow{2}{*}{\textbf{Method}} &
\textbf{Classification} &
\textbf{Generation} &
\multicolumn{2}{c}{\textbf{Evidence}} &
\multicolumn{2}{c}{\textbf{Description}} \\
\cmidrule(lr){3-3} \cmidrule(lr){4-4} \cmidrule(lr){5-6} \cmidrule(lr){7-8}
 & & \textbf{Acc / JClsAcc\%}
   & \textbf{JSON Valid\%}
   & \textbf{ROUGE-L} & \textbf{BERTScore}
   & \textbf{ROUGE-L} & \textbf{BERTScore} \\
\midrule
\multirow{4}{*}{ISCX-Tor-2016}
  & NetMamba          & \textbf{0.9961} & --       & --     & --     & --     & --     \\
  & Zero-shot LLM     & 0.0003 & 100.00\% & 0.1247 & 0.8322 & 0.1164 & 0.8469 \\
  & Vanilla        & 0.7092 & 100.00\% & 0.6002 & 0.9217 & 0.5831 & 0.9266 \\
  & \textbf{mmTraffic (Ours)} & 0.9331 & \textbf{100.00\%} & \textbf{0.8192} & \textbf{0.9641} & \textbf{0.7751} & \textbf{0.9481} \\
\midrule
\multirow{4}{*}{ISCXVPN2016}
  & NetMamba          & \textbf{0.9917} & --       & --     & --     & --     & --     \\
  & Zero-shot LLM     & 0.0004 & 100.00\% & 0.1121 & 0.8290 & 0.1252 & 0.8545 \\
  & Vanilla        & 0.2987 & 100.00\% & 0.3597 & 0.8881 & 0.2020 & 0.8679 \\
  & \textbf{mmTraffic (Ours)} & 0.9902 & \textbf{100.00\%} & \textbf{0.8436} & \textbf{0.9686} & \textbf{0.6975} & \textbf{0.9419} \\
\midrule
\multirow{4}{*}{CSTNet-TLS1.3}
  & NetMamba          & \textbf{0.8474} & --       & --     & --     & --     & --     \\
  & Zero-shot LLM     & 0.0000 & 100.00\% & 0.2675 & 0.8780 & 0.1399 & 0.8492 \\
  & Vanilla        & 0.0148 & 100.00\% & 0.5224 & 0.9242 & 0.4346 & 0.9041 \\
  & \textbf{mmTraffic (Ours)} & 0.6448 & \textbf{100.00\%} & \textbf{0.7188} & \textbf{0.9538} & \textbf{0.8007} & \textbf{0.9710} \\
\bottomrule
\end{tabular}%
}
\end{table*}

\subsubsection{Implementation Details}

The traffic encoder $\mathrm{T}_{\theta}$ is instantiated with NetMamba~\cite{NetMamba}. Unlike previous decoupled methods, $\mathrm{T}_{\theta}$ is completely unfrozen and fully fine-tuned to actively capture language-aligned semantic representations. The alignment connector $\mathrm{C}_{\omega}$ is implemented as a two-layer MLP with GELU activation (\texttt{mlp2x\_gelu}) and is also fully fine-tuned, alongside the newly introduced auxiliary classification head $\mathrm{A}_{\kappa}$. The cognitive module $\mathrm{G}_{\phi}$ is instantiated with Qwen3-1.7B~\cite{qwen3} and adapted via Low-Rank Adaptation (LoRA)~\cite{lora}. LoRA is applied to all attention and feed-forward projection modules (specifically \texttt{<q\_proj>}, \texttt{<k\_proj>}, \texttt{<v\_proj>}, \texttt{<o\_proj>}, \texttt{<gate\_proj>}, \texttt{<up\_proj>}, and \texttt{<down\_proj>}), with an increased rank $r = 32$, scaling factor $\alpha = 64$, and a dropout rate of 0.1. Thus, the parameters of $\mathrm{T}_{\theta}$, $\mathrm{C}_{\omega}$, $\mathrm{A}_{\kappa}$, and the LoRA modules of $\mathrm{G}_{\phi}$ are jointly updated during the end-to-end training.

For the multi-task optimization objectives, the balancing weight for the auxiliary classification loss is set to $\lambda = 0.3$. For the semantic-priority guided generation loss, we set the categorical boundary threshold to $M = 15$ and the boost weight factor to $\gamma = 5.0$, firmly anchoring the text generation to the physical traffic identity.

All models are trained for 10 epochs using the AdamW~\cite{AdamW} optimizer with a peak learning rate of $\eta = 5 \times 10^{-5}$, a weight decay of 0.01, a linear warmup~\cite{Linear_warmup} over the first 10\% of training steps, and a gradient clipping threshold of 1.0. We utilize BFloat16~\cite{mixed_BFloat16} mixed-precision and distributed training via DeepSpeed ZeRO-2~\cite{zero}. The per-device batch size is set to 3 with a gradient accumulation of 8 steps, yielding a global batch size of $3 \times 8 \times 5 = 120$ across 5 NVIDIA A800 GPUs.

\subsubsection{Evaluation Metrics}
\label{sec:metrics}

We evaluate \textbf{mmTraffic} from two complementary perspectives: \textit{traffic classification} and \textit{forensic report generation}. The evaluation metrics include:
\begin{itemize}
\item \textbf{Classification Metrics}. We report \textit{Accuracy} to assess the traffic identification performance of the perception module, measured as the proportion of correctly classified samples over the full test set. We additionally report \textit{JSON Validity Rate} (JSON Valid\%), the proportion of model outputs that can be successfully parsed as a valid JSON object containing all required fields, which reflects whether the model has learned the structured output format.

\item \textbf{Text Generation Metrics}. To assess the quality of the generated evidence and description fields, we adopt two complementary metrics:
    \textit{ROUGE-L}~\cite{rouge-l} measures the $F_1$ score of the longest common subsequence (LCS) between the generated and reference texts, capturing lexical overlap and word-order consistency.
     \textit{BERTScore}~\cite{bertscore} computes token-level semantic similarity between generated and reference texts using contextual embeddings from \texttt{roberta-large} (\texttt{num\_layers=17}), loaded from a local checkpoint to ensure reproducibility. We report the macro-averaged BERTScore F$_1$, computed as the arithmetic mean of per-sample $F_1$ scores (harmonic mean of token-level precision and recall) over the full test set. Compared to ROUGE-L, BERTScore is more robust to lexical paraphrasing and stylistic variation between the ground-truth anchor (Claude Opus) and the prediction model (Qwen3-1.7B)~\cite{qwen3}, and thus a reliable indicator of semantic fidelity when surface-level wording differs.

\item \textbf{Structural Consistency Metrics.} Beyond lexical and semantic similarity, we evaluate the internal quality of generated reports using three \textit{reference-free} metrics, which are visualized in the radar charts, computed solely over the model's generated output, requiring no ground-truth text for evaluation. Let $N$ denote the number of samples with valid JSON predictions, and let $e_i$, $d_i$ denote the predicted evidence and description fields of the $i$-th report, with $c_i = [e_i ; d_i]$ denoting their concatenation.
    \textit{Evidence-Trait Consistency} (ETC) measures whether the generated evidence text is semantically coherent with the predicted byte-level trait values, verifying that the model's reasoning is grounded in the actual traffic features rather than generating plausible-sounding but ungrounded observations. It is computed as:
    \begin{equation}
         \text{ETC} = \frac{1}{N} \sum_{i=1}^{N} \mathbf{1}[\text{KW}(\mathcal{T}_i) \cap e_i \neq \emptyset]
    \end{equation}
    where $\mathbf{1}[\cdot]$ is the indicator function that equals 1 if the condition holds and 0 otherwise, $\text{KW}(\mathcal{T}_i)$ denotes the union of keyword sets of all predicted traits, and $e_i$ denotes the tokens of the predicted evidence text. 

    \textit{Quantitative Claim Rate} (QCR) measures the proportion of reports containing at least one concrete numerical observation, such as byte counts, entropy values, or explicit ordinal descriptors. A high QCR indicates that the model produces specific, verifiable reports rather than vague qualitative descriptions. It is computed as:
    \begin{equation}
        \text{QCR} = \frac{1}{N} \sum_{i=1}^{N} \mathbf{1}[\text{HasQuant}(c_i)]
    \end{equation}
    where $\text{HasQuant}(\cdot)$ is true if $c_i$ contains any percentage, byte quantity, multi-digit number, or ordinal descriptor (\texttt{high}/\texttt{mid}/\texttt{low}), or the keyword \texttt{ratio}.

   \textit{Protocol Mention Rate} (PMR) measures the proportion of reports that explicitly reference at least one network protocol by name or identifier (e.g., TCP, TLS, HTTP, QUIC). Protocol attribution is a fundamental requirement for reports, and a high PMR confirms the model reliably grounds its analysis in appropriate protocol context. 
    \begin{equation}
        \text{PMR} = \frac{1}{N} \sum_{i=1}^{N} \mathbf{1}[\mathcal{P} \cap c_i \neq \emptyset]
    \end{equation}
    where $\mathcal{P}$ is a predefined set of protocol keywords (e.g., TCP, TLS, HTTP).
    
We note that reference-based metrics such as ROUGE-L and BERTScore exhibit insensitivity to classification correctness in fine-grained settings, as the ground-truth descriptions across categories share substantial protocol-level vocabulary. This motivates the introduction of reference-free structural consistency metrics.
\end{itemize}

\begin{table*}[t]
\centering
\caption{Results on CrossPlatform-iOS, CrossPlatform-Android, and USTC-TFC-2016. \textbf{NetMamba} is the unimodal classifier (``--'' means no generation); \textbf{Zero-shot LLM} feeds features directly to LLM without tuning; \textbf{Vanilla} freezes the encoder without auxiliary constraints. 
\textbf{Bold} indicates the best per column within each dataset group.}
\label{tab:main_results_2}
\resizebox{\linewidth}{!}{%
\begin{tabular}{l l c c cc cc}
\toprule
\multirow{2}{*}{\textbf{Dataset}} &
\multirow{2}{*}{\textbf{Method}} &
\textbf{Classification} &
\textbf{Generation} &
\multicolumn{2}{c}{\textbf{Evidence}} &
\multicolumn{2}{c}{\textbf{Description}} \\
\cmidrule(lr){3-3} \cmidrule(lr){4-4} \cmidrule(lr){5-6} \cmidrule(lr){7-8}
 & & \textbf{Acc / JClsAcc\%}
   & \textbf{JSON Valid\%}
   & \textbf{ROUGE-L} & \textbf{BERTScore}
   & \textbf{ROUGE-L} & \textbf{BERTScore} \\
\midrule
\multirow{4}{*}{CrossPlatform-iOS}
  & NetMamba          & \textbf{0.9060} & --       & --     & --     & --     & --     \\
  & Zero-shot LLM     & 0.0000 & 100.00\% & 0.1962 & 0.8509 & 0.1268 & 0.8503 \\
  & Vanilla        & 0.0058 & 100.00\% & 0.2218 & 0.8591 & 0.1255 & 0.8535 \\
  & \textbf{mmTraffic (Ours)} & 0.8865 & \textbf{100.00\%} & \textbf{0.6880} & \textbf{0.9387} & \textbf{0.5972} & \textbf{0.9283} \\
\midrule
\multirow{4}{*}{CrossPlatform-Android}
  & NetMamba          & \textbf{0.9104} & --       & --     & --     & --     & --     \\
  & Zero-shot LLM     & 0.0000 & 100.00\% & 0.0000 & 0.0000 & 0.0000 & 0.8405 \\
  & Vanilla        & 0.0027 & 100.00\% & 0.2107 & 0.8661 & 0.1283 & 0.8542 \\
  & \textbf{mmTraffic (Ours)} & 0.8654 & \textbf{100.00\%} & \textbf{0.5482} & \textbf{0.9060} & \textbf{0.5605} & \textbf{0.9299} \\
\midrule
\multirow{4}{*}{USTC-TFC-2016}
  & NetMamba          & \textbf{0.9887} & --       & --     & --     & --     & --     \\
  & Zero-shot LLM     & 0.0000 & 100.00\% & 0.1386 & 0.8377 & 0.1365 & 0.8536 \\
  & Vanilla        & 0.7002 & 100.00\% & 0.6383 & 0.9272 & 0.5447 & 0.9163 \\
  & \textbf{mmTraffic (Ours)} & 0.8624 & \textbf{100.00\%} & \textbf{0.8853} & \textbf{0.9769} & \textbf{0.7714} & \textbf{0.9527} \\
\bottomrule
\end{tabular}%
}
\end{table*}


\subsection{Main Results}
\label{sec:main_results}
Tables~\ref{tab:main_results_1} and~\ref{tab:main_results_2} present the comparison results with baselines. We report the linear head Accuracy (Acc) for NetMamba, while the JSON Classification Accuracy (JClsAcc\%) extracted directly from the generated natural language reports for generative models (Zero-shot LLM, Vanilla, and ours).

\textbf{Evaluation of Classification Performance.} NetMamba sets the upper-bound baseline for specialized unimodal classification. As observed, the \textit{Zero-shot LLM} and \textit{Vanilla} paradigms suffer a catastrophic drop in classification capability, failing almost entirely on datasets like CSTNet-TLS1.3 (0.0148) and CrossPlatform-iOS (0.0058). This collapse demonstrates that without joint optimization, the semantic gap between physical bytes and the lexical space is insurmountable. In contrast, our proposed \textbf{mmTraffic} successfully bridges this gap. By unfreezing the encoder and applying the auxiliary classification head, mmTraffic recovers robust JSON classification performance (e.g., reaching 0.9902 on ISCXVPN2016 and 0.8865 on CrossPlatform-iOS). While there is a slight inherent \textit{alignment tax} compared to the pure linear classifier (e.g., 0.9887 for NetMamba vs. 0.8624 for mmTraffic on USTC-TFC-2016) due to the complexity of autoregressive text generation, it remains highly competitive and overwhelmingly surpasses standard multimodal baselines (i.e., zero-shot LLM and Vanilla).

\textbf{Evaluation of Generation Quality.} Generating human-readable and evidence-grounded reports is the core objective. Unimodal classifiers like NetMamba are fundamentally incapable of generating text. The \textit{Vanilla} manages to produce fluent text but generates severe hallucinations due to its inability to accurately identify the underlying traffic. Conversely, \textbf{mmTraffic} achieves a JSON validity rate of 100\% across all datasets, confirming that the LLM has fully mastered the structured output format. On evidence and description generation, mmTraffic demonstrates overwhelming superiority. For example, it achieves an Evidence ROUGE-L of 0.8436 on ISCXVPN2016 and a Description BERTScore of 0.9710 on CSTNet-TLS1.3. Across all six datasets, the BERTScore consistently remains above 0.90, proving that the generated evidence and behavioral descriptions maintain rigorous semantic alignment with the ground-truth expert annotations.

\begin{figure}[t]
\centering
\includegraphics[width=\linewidth]{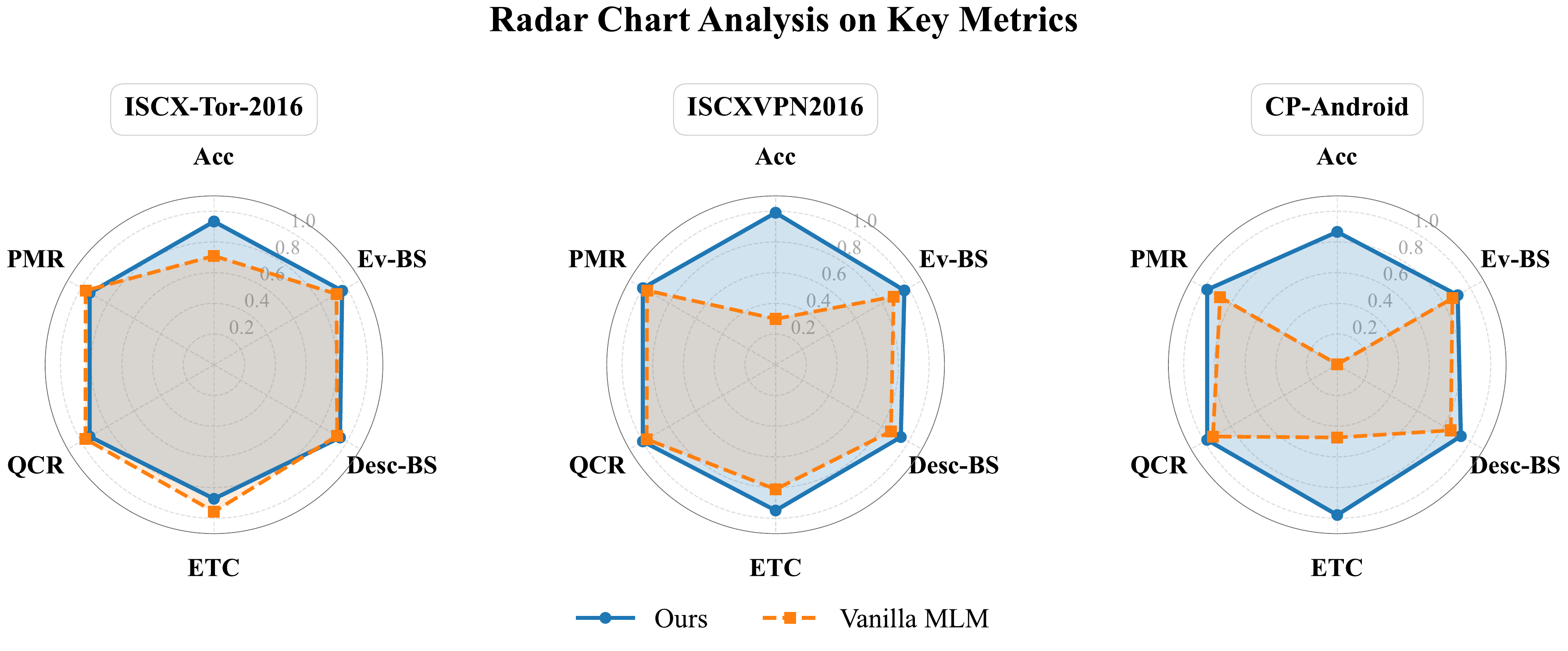}
\caption{Analysis on Structural Consistency Metrics. The semantic-priority constraints in mmTraffic ensure high logical rigor.}
\label{fig:radar}
\end{figure}

\textbf{Evaluation of Structural Consistency.} A critical observation in multimodal traffic analysis is the "fluency trap": models like \textit{Vanilla MLM}  might achieve moderate text generation metrics (e.g., Ev-BS and Desc-BS) by memorizing common vocabulary, even when their classification accuracy (Acc) completely collapses—as starkly visible in the ISCXVPN2016 and CP-Android radar charts. This reveals a fundamental limitation of reference-based metrics in forensics—lexical overlap does not guarantee factual correctness. The radar charts in Figure~\ref{fig:radar} provide a comprehensive picture of logical rigorousness across six key dimensions. While \textit{Vanilla MLM} exhibits severely distorted performance profiles , our proposed \textbf{mmTraffic} maintains a robust, near-symmetrical shape that pushes towards the outer boundaries (1.0)  across all evaluated datasets. Driven by our semantic-priority guided generation mechanism, mmTraffic ensures that its high structural consistency (ETC, QCR, PMR) is strictly anchored in correct traffic identification, effectively eliminating the multimodal hallucinations prevalent in unconstrained architectures.

\begin{table}[t]
\centering
\caption{ISCX-Tor-2016 for sample \#6227: mmTraffic correctly identifies the class. Evidence and description fields are abbreviated.
\textcolor{green!60!black}{\textbf{Green}} means correctly interpreted key content.}
\label{tab:tor_case_a}
\newcolumntype{G}{>{\columncolor{green!6}}p{5.5cm}}
\newcolumntype{H}{>{\columncolor{gray!6}}p{5.5cm}}
\resizebox{\linewidth}{!}{%
\begin{tabular}{l G H}
\toprule
\textbf{Field} & \textbf{mmTraffic (Ours)} & \textbf{Ground Truth} \\
\midrule
\rowcolor{gray!15}
\textbf{Class}
  & \textcolor{green!60!black}{\textbf{CHAT}} \checkmark
  & CHAT \\
\midrule
\rowcolor{blue!5}
\textbf{Traits}
  & has\_tls: \textcolor{green!60!black}{\textbf{\texttt{true}}} \newline
    has\_http: \textcolor{green!60!black}{\textbf{\texttt{false}}} \newline
    ascii: \texttt{low} \newline
    entropy: \textcolor{green!60!black}{\textbf{\texttt{mid}}} \newline
    zero\_pad: \textcolor{green!60!black}{\textbf{\texttt{mid}}}
  & has\_tls: \texttt{true} \newline
    has\_http: \texttt{false} \newline
    ascii: \texttt{mid} \newline
    entropy: \texttt{mid} \newline
    zero\_pad: \texttt{mid} \\
\midrule
\rowcolor{blue!5}
\textbf{Evidence}
  & (1) \textcolor{green!60!black}{\textbf{TLS record header detected}}, indicating
    an encrypted TLS session. \newline
    (2) \textcolor{green!60!black}{\textbf{TCP dominant (100\%)}}, consistent with
    Tor CHAT profile. \newline
    (3) Unlike Tor VOIP (codec-driven timing), \textcolor{green!60!black}{\textbf{Tor CHAT
    has irregular human-conversation-driven timing}}. \ldots
  & (1) TLS record header detected, indicating an encrypted
    TLS session. \newline
    (2) TLS dominant (100\%), consistent with Tor CHAT
    profile. \newline
    (3) Unlike Tor VOIP (codec-driven timing), Tor CHAT
    has irregular human-conversation-driven timing. \ldots \\
\midrule
\rowcolor{yellow!8}
\textbf{Description}
  & TLS record headers present, indicating encrypted transport.
    \textbf{\textcolor{green!60!black}{Instant messaging and chat
    protocols (AIM, ICQ, Facebook Chat, Hangouts, Skype Chat)}}
    tunneled through Tor onion routing.
    Dominant protocol: TCP (100\%). \ldots
  & TLS record headers present, indicating encrypted transport.
    \textbf{\textcolor{gray!60!black}{Instant messaging and chat
    protocols (AIM, ICQ, Facebook Chat, Hangouts, Skype Chat)}}
    tunneled through Tor onion routing.
    Dominant protocol: TLS (100\%). \\
\midrule
\rowcolor{gray!5}
\textbf{Notes}
  & Chat over Tor hides who is communicating with whom; message
    content may still be visible at exit nodes if not
    end-to-end encrypted.
  & Chat over Tor provides metadata anonymity; message content
    may still be visible at exit nodes if not end-to-end
    encrypted. \\
\bottomrule
\end{tabular}%
}
\end{table}

\begin{table}[t]
\centering
\caption{CSTNet-TLS1.3 for sample \#350: mmTraffic correctly identifies the class. Evidence and
description fields are abbreviated. \textcolor{green!60!black}{\textbf{Green}} means correctly interpreted key content.}
\label{tab:cstnet_case_a}
\newcolumntype{G}{>{\columncolor{green!6}}p{5.5cm}}
\newcolumntype{H}{>{\columncolor{gray!6}}p{5.5cm}}
\resizebox{\linewidth}{!}{%
\begin{tabular}{l G H}
\toprule
\textbf{Field} & \textbf{mmTraffic (Ours)} & \textbf{Ground Truth} \\
\midrule
\rowcolor{gray!15}
\textbf{Class}
  & \textcolor{green!60!black}{\textbf{Steam}} \checkmark
  & Steam \\
\midrule
\rowcolor{blue!5}
\textbf{Traits}
  & has\_tls: \textcolor{green!60!black}{\textbf{\texttt{true}}} \newline
    has\_http: \textcolor{green!60!black}{\textbf{\texttt{false}}} \newline
    ascii: \texttt{mid} \newline
    entropy: \textcolor{green!60!black}{\textbf{\texttt{high}}} \newline
    zero\_pad: \textcolor{green!60!black}{\textbf{\texttt{low}}}
  & has\_tls: \texttt{true} \newline
    has\_http: \texttt{false} \newline
    ascii: \texttt{low} \newline
    entropy: \texttt{high} \newline
    zero\_pad: \texttt{low} \\
\midrule
\rowcolor{blue!5}
\textbf{Evidence}
  & (1) \textcolor{green!60!black}{\textbf{TLS record header detected}},
    indicating an encrypted TLS session. \newline
    (2) \textcolor{green!60!black}{\textbf{High Shannon entropy}}: encrypted
    or compressed payload. \newline
    (3) \textcolor{green!60!black}{\textbf{Low zero-padding}}: sustained
    bulk data transfer. \ldots
  & (1) TLS record header detected, indicating an encrypted
    TLS session. \newline
    (2) Low ASCII with high entropy: encrypted payload. \newline
    (3) Low zero-padding: sustained bulk data transfer. \newline
    (4) TCP dominant (84.2\%). \ldots \\
\midrule
\rowcolor{yellow!8}
\textbf{Description}
  & TLS record headers alongside high-entropy payload confirm
    encrypted communication.
    \textbf{\textcolor{green!60!black}{Valve Steam gaming platform}}
    for game downloads, updates, and multiplayer services
    over TLS 1.3. Large avg. packet (858 bytes). \ldots
  & TLS record headers alongside high-entropy payload confirm
    encrypted communication.
    \textbf{\textcolor{gray!60!black}{Valve Steam gaming platform}}
    for game downloads, updates, and multiplayer services
    over TLS 1.3. TCP dominant (84.2\%). \ldots \\
\midrule
\rowcolor{gray!5}
\textbf{Notes}
  & Game distribution platform; monitor for Steam account
    phishing; Workshop mods can contain malicious code.
  & Game distribution platform; game downloads are very large;
    monitor for Steam account phishing. \\
\bottomrule
\end{tabular}%
}
\end{table}

\begin{table}[t]
\centering
\caption{USTC-TFC-2016 for sample \#11786: mmTraffic correctly identifies the class. Evidence and
description fields are abbreviated.
\textcolor{green!60!black}{\textbf{Green}} means correctly interpreted key content.}
\label{tab:ustc_case_a}
\newcolumntype{G}{>{\columncolor{green!6}}p{5.5cm}}
\newcolumntype{H}{>{\columncolor{gray!6}}p{5.5cm}}
\resizebox{\linewidth}{!}{%
\begin{tabular}{l G H}
\toprule
\textbf{Field} & \textbf{mmTraffic(Ours)} & \textbf{Ground Truth} \\
\midrule
\rowcolor{gray!15}
\textbf{Class}
  & \textcolor{green!60!black}{\textbf{Outlook}} \checkmark
  & Outlook \\
\midrule
\rowcolor{blue!5}
\textbf{Traits}
  & has\_tls: \textcolor{green!60!black}{\textbf{\texttt{true}}} \newline
    has\_http: \textcolor{green!60!black}{\textbf{\texttt{false}}} \newline
    ascii: \textcolor{green!60!black}{\textbf{\texttt{low}}} \newline
    entropy: \texttt{mid} \newline
    zero\_pad: \texttt{mid}
  & has\_tls: \texttt{true} \newline
    has\_http: \texttt{false} \newline
    ascii: \texttt{low} \newline
    entropy: \texttt{high} \newline
    zero\_pad: \texttt{low} \\
\midrule
\rowcolor{blue!5}
\textbf{Evidence}
  & (1) \textcolor{green!60!black}{\textbf{TLS record header detected}},
    indicating an encrypted TLS session. \newline
    (2) \textcolor{green!60!black}{\textbf{TLS dominant (100\%)}}, consistent
    with Outlook profile. \newline
    (3) Unlike Gmail (HTTP/2 and QUIC), \textcolor{green!60!black}{\textbf{Outlook
    uses MAPI protocol}}. \ldots
  & (1) TLS record header detected, indicating an encrypted
    TLS session. \newline
    (2) Low ASCII with high entropy: encrypted payload. \newline
    (3) Low zero-padding: sustained bulk data transfer. \newline
    (4) TLS dominant (100\%). \ldots \\
\midrule
\rowcolor{yellow!8}
\textbf{Description}
  & TLS record headers present, indicating encrypted transport.
    \textbf{\textcolor{green!60!black}{Microsoft Outlook email
    client using HTTPS and MAPI}}.
    Dominant protocol: TLS (100\%). \ldots
  & TLS record headers alongside high-entropy payload confirm
    encrypted communication.
    \textbf{\textcolor{gray!60!black}{Microsoft Outlook email
    client using HTTPS and MAPI}}.
    Dominant protocol: TLS (100\%). \\
\midrule
\rowcolor{gray!5}
\textbf{Notes}
  & Generally benign; verify connections go to Microsoft ASN;
    anomalous sync volumes may indicate data exfiltration.
  & Generally benign; verify connections go to Microsoft ASN;
    anomalous sync volumes or off-hours activity may indicate
    data exfiltration via email. \\
\bottomrule
\end{tabular}%
}
\end{table}

\begin{table*}[t]
\centering
\tiny
\caption{ISCX-Tor-2016 for sample \#15750: mmTraffic correctly identifies the VIDEO class while \textit{Vanilla VLM} misclassifies it as BROWSING.
Evidence and description fields are abbreviated.
\textcolor{green!60!black}{\textbf{Green}} and \textcolor{red!70!black}{Red} mean correct and
wrong interpretations.}
\label{tab:tor_case_b}
\newcolumntype{D}{>{\columncolor{green!6}}p{5.0cm}}
\newcolumntype{E}{>{\columncolor{red!6}}p{5.0cm}}
\newcolumntype{F}{>{\columncolor{gray!6}}p{5.0cm}}
\resizebox{\linewidth}{!}{%
\begin{tabular}{l D E F}
\toprule
\textbf{Field}
  & \textbf{mmTraffic (Ours)}
  & \textbf{Vanilla VLM}
  & \textbf{Ground Truth} \\
\midrule
\rowcolor{gray!15}
\textbf{Class}
  & \textcolor{green!60!black}{\textbf{VIDEO}} \checkmark
  & \textcolor{red!70!black}{\textbf{BROWSING}} \XSolidBrush 
  & VIDEO \\
\midrule
\rowcolor{blue!5}
\textbf{Evidence}
  & (1) \textcolor{green!60!black}{\textbf{Video streaming services (YouTube, Vimeo)}}
    tunneled via Tor onion routing. \newline
    (2) High throughput (49.8 MB/s) indicates active data exchange. \newline
    (3) \textcolor{green!60!black}{\textbf{Sustained download dominant over upload}},
    consistent with video consumption. \ldots
  & (1) TLS record header detected, indicating encrypted session. \newline
    (2) \textcolor{red!70!black}{\textbf{High zero-padding: short flow with small
    payload}}. \newline
    (3) TCP dominant (100\%), \textcolor{red!70!black}{\textbf{consistent with
    HTTPS browsing profile}}. \ldots
  & (1) TCP dominant (100\%), consistent with VIDEO profile. \newline
    (2) Unlike Tor AUDIO (moderate throughput), VIDEO requires
    significantly higher bandwidth. \newline
    (3) Unlike Tor BROWSING (bursty), VIDEO maintains sustained
    high-throughput download. \ldots \\
\midrule
\rowcolor{yellow!8}
\textbf{Description}
  & Mixed entropy and ASCII across packet slices.
    \textbf{\textcolor{green!60!black}{Video streaming services
    (youtube.com, vimeo.com)}} tunneled via Tor onion routing.
    TCP dominant. High throughput (498 KB/s). \ldots
  & TLS record headers indicate encrypted transport.
    \textbf{\textcolor{red!70!black}{HTTPS/SSL web browsing
    tunneled through Tor}} hides both location and activities.
    High zero-padding suggests background activity. \ldots
  & Mixed entropy and ASCII across packet slices.
    \textbf{\textcolor{gray!60!black}{Video streaming services
    (YouTube HTML5, YouTube Flash, Vimeo)}} tunneled through
    Tor. Dominant protocol: TCP (100\%). \\
\bottomrule
\end{tabular}%
}
\end{table*}

\begin{table*}[t]
\centering
\tiny
\caption{CSTNet-TLS1.3 for sample \#4728: mmTraffic correctly identifies the \texttt{Adobe} while \textit{Vanilla VLM} misclassifies it as \texttt{baidu.com}. Under TLS 1.3, byte-level features are nearly identical across all websites.
\textcolor{green!60!black}{\textbf{Green}} and \textcolor{red!70!black}{Red} mean correct and
wrong interpretations.}
\label{tab:cstnet_case_b}
\newcolumntype{D}{>{\columncolor{green!6}}p{5.0cm}}
\newcolumntype{E}{>{\columncolor{red!6}}p{5.0cm}}
\newcolumntype{F}{>{\columncolor{gray!6}}p{5.0cm}}
\resizebox{\linewidth}{!}{%
\begin{tabular}{l D E F}
\toprule
\textbf{Field}
  & \textbf{mmTraffic (Ours)}
  & \textbf{Vanilla VLM}
  & \textbf{Ground Truth} \\
\midrule
\rowcolor{gray!15}
\textbf{Class}
  & \textcolor{green!60!black}{\textbf{Adobe}} \checkmark
  & \textcolor{red!70!black}{\textbf{baidu.com}} \XSolidBrush 
  & Adobe \\
\midrule
\rowcolor{blue!5}
\textbf{Evidence}
  & (1) \textcolor{green!60!black}{\textbf{TLS record header detected}}. \newline
    (2) \textcolor{green!60!black}{\textbf{High Shannon entropy}}: encrypted
    or binary payload. \newline
    (3) \textcolor{green!60!black}{\textbf{Low zero-padding}}: sustained bulk
    data transfer. \ldots
  & (1) TLS record header detected. \newline
    (2) High zero-padding: short flow with small payload. \newline
    (3) TCP dominant (68.5\%), \textcolor{red!70!black}{\textbf{consistent
    with Baidu's search and CDN infrastructure}}. \ldots
  & (1) TLS record header detected. \newline
    (2) Low ASCII with high entropy: encrypted payload. \newline
    (3) TLS dominant (55.6\%), consistent with Adobe
    profile. \ldots \\
\midrule
\rowcolor{yellow!8}
\textbf{Description}
  & TLS record headers alongside high-entropy payload confirm
    encrypted communication.
    \textbf{\textcolor{green!60!black}{Adobe creative software,
    cloud services, and document management platform}}
    over TLS 1.3. \ldots
  & TLS record headers indicate active encrypted communication.
    \textbf{\textcolor{red!70!black}{Large volume of bulk
    control-plane packets over TCP. Baidu}} search and CDN
    infrastructure. \ldots
  & TLS record headers alongside high-entropy payload confirm
    encrypted communication.
    \textbf{\textcolor{gray!60!black}{Adobe creative software,
    cloud services, and document management platform}}
    over TLS 1.3. \ldots \\
\bottomrule
\end{tabular}%
}
\end{table*}

\begin{table*}[t]
\centering
\tiny
\caption{USTC-TFC-2016 for sample \#10195: mmTraffic correctly identifies the Geodo malware flow while \textit{Vanilla VLM} misclassifies it as Htbot. Both are HTTP-based botnets with similar byte signatures.
\textcolor{green!60!black}{\textbf{Green}} and \textcolor{red!70!black}{Red} mean correct and
wrong interpretations.}
\label{tab:ustc_case_b}
\newcolumntype{D}{>{\columncolor{green!6}}p{5.0cm}}
\newcolumntype{E}{>{\columncolor{red!6}}p{5.0cm}}
\newcolumntype{F}{>{\columncolor{gray!6}}p{5.0cm}}
\resizebox{\linewidth}{!}{%
\begin{tabular}{l D E F}
\toprule
\textbf{Field}
  & \textbf{mmTraffic (Ours)}
  & \textbf{Vanilla VLM}
  & \textbf{Ground Truth} \\
\midrule
\rowcolor{gray!15}
\textbf{Class}
  & \textcolor{green!60!black}{\textbf{Geodo}} \checkmark
  & \textcolor{red!70!black}{\textbf{Htbot}} \XSolidBrush
  & Geodo \\
\midrule
\rowcolor{blue!5}
\textbf{Evidence}
  & (1) \textcolor{green!60!black}{\textbf{High Shannon entropy}}: encrypted
    or binary payload. \newline
    (2) \textcolor{green!60!black}{\textbf{Low zero-padding}}: sustained bulk
    data transfer. \newline
    (3) TCP dominant (100\%). \newline
    (4) Unlike Zeus (banking only), \textcolor{green!60!black}{\textbf{Geodo
    also performs spam relay}}. 
  & (1) HTTP tokens (GET/POST) indicate plaintext HTTP
    requests. \newline
    (2) High ASCII: plaintext application-layer content. \newline
    (3) Low zero-padding: sustained bulk data transfer. \ldots
  & (1) High zero-padding: short flow, small payload. \newline
    (2) TCP dominant (100\%). \newline
    (3) Unlike Zeus (banking only), Geodo also performs
    spam relay. \ldots \\
\midrule
\rowcolor{yellow!8}
\textbf{Description}
  & Mixed entropy and ASCII characteristics.
    \textbf{\textcolor{green!60!black}{Geodo/Emotet banking trojan
    and email spam botnet}}. Dominant protocol: TCP (100\%). \ldots
  & HTTP method tokens and readable ASCII confirm plaintext
    \textbf{\textcolor{red!70!black}{botnet-driven web scraping
    or C\&C command execution (Htbot)}}. Low throughput suggests
    sparse traffic. \ldots
  & Mixed entropy and ASCII characteristics.
    \textbf{\textcolor{gray!60!black}{Geodo/Emotet banking trojan
    and email spam botnet}}. Dominant protocol: TCP (100\%). \\
\bottomrule
\end{tabular}%
}
\end{table*}

\begin{figure}[t]
  \centering
  \includegraphics[width=\linewidth]{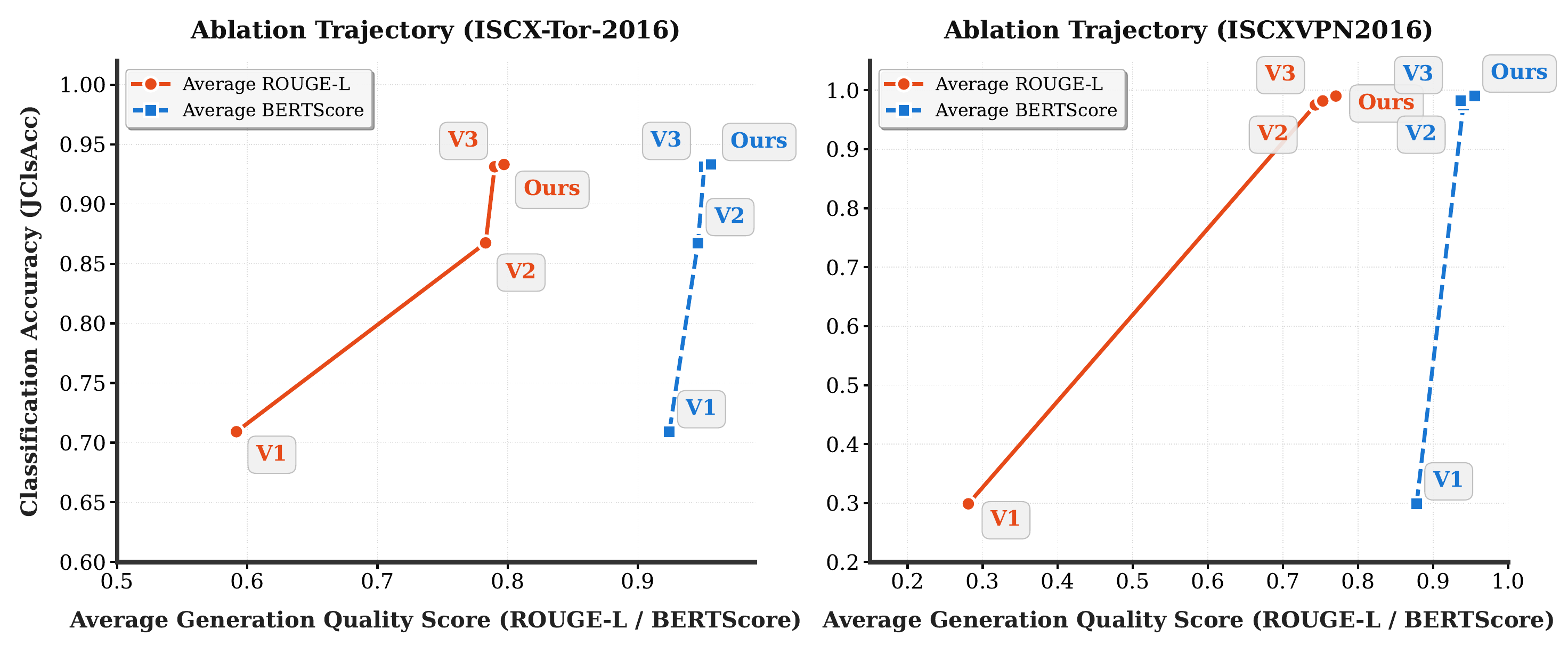}
  \caption{Ablation analysis on ISCX-Tor-2016 and ISCXVPN2016, with respect to the classification and generation metrics for four variants from V1 to V4.}
  \label{fig:ablation_profile}
\end{figure}

\subsection{Ablation Study}
\label{sec:ablation}

To isolate the contribution of each proposed component in mmTraffic, we conduct an ablation study across two distinct domains: ISCX-Tor-2016 and ISCXVPN2016. As illustrated in Figure~\ref{fig:ablation_profile}, we systematically evaluate four configurations: (1) \textbf{V1 (Vanilla MLLM)}: freezing the NetMamba encoder and relying solely on the standard Negative Log-Likelihood (NLL) loss; (2) \textbf{V2 (+ Unfrozen)}: unfreezing the traffic encoder for end-to-end joint optimization; (3) \textbf{V3 (+ Auxiliary Head)}: introducing the auxiliary classification head ($\mathcal{L}_{\mathrm{aux}}$) to the latent space; and (4) \textbf{V4 (mmTraffic Full)}: incorporating the semantic-priority guided generation mechanism ($\mathcal{L}_{\mathrm{gen}}$).

\textbf{Breaking the Modality Barrier via Joint Optimization.} 
The initial transition from V1 to V2 in Figure~\ref{fig:ablation_profile} reveals the fundamental bottleneck of cross-modal traffic analysis. In the \textit{Vanilla MLLM} (V1) setting, the framework suffers a catastrophic failure on ISCXVPN2016 (accuracy at 0.2987) and struggles at 0.7092 on ISCX-Tor-2016. Because raw cryptographic traffic sequences lack the natural lexical alignments found in visual-text data, a frozen encoder fundamentally fails to project these non-semantic bytes into the LLM's sophisticated cognitive space. Unfreezing the encoder (\textit{+ Unfrozen}) opens the gradient bottleneck, allowing the perceptual backbone to dynamically adapt its feature extraction guided by the LLM's generation objective. This mechanistic bridge yields a massive concurrent leap in both text fidelity (e.g., Average ROUGE-L on Tor rises from 0.59 to 0.78) and classification accuracy (0.8674 on Tor and 0.9751 on VPN).

\textbf{Shaping the Latent Space with Auxiliary Constraints.}
While unfreezing the encoder bridges the modality gap, relying exclusively on the LLM's autoregressive text-generation loss provides weak and implicit supervision, which is insufficient to disentangle highly overlapping encrypted traffic patterns. The critical inflection point occurs in V3 with the introduction of the auxiliary classification head. By directly penalizing misclassifications at the feature bottleneck, $\mathcal{L}_{\mathrm{aux}}$ explicitly reshapes the continuous latent space. It forces the encoder to establish hard, discriminative boundaries before the features ever reach the LLM. This explicit concept anchoring effectively resolves perceptual ambiguity, propelling the classification accuracy to 0.9312 on ISCX-Tor-2016 and 0.9819 on ISCXVPN2016, while maintaining strong generative performance.

\textbf{Synergistic Grounding via Semantic-Priority Generation.}
The final transition to \textbf{mmTraffic (Full)} demonstrates that our semantic-priority generation loss ($\mathcal{L}_{\mathrm{gen}}$) is not merely a linguistic constraint, but a mechanism for cognitive synergy. In standard unconstrained generation (V3), all tokens are treated equally, leaving the model susceptible to generating fluent but ungrounded priors. By dynamically assigning a heavy penalty weight to the categorical prefix tokens, mmTraffic forces the LLM to commit strictly to a physical traffic identity first. Remarkably, rather than acting as a restrictive trade-off, this strong semantic grounding mechanism stabilizes the reasoning chain, pushing the final classification accuracy to its peak across both domains (0.9331 on Tor and 0.9902 on VPN) and maximizing the structural alignment of the generated evidence (Average BERTScore reaches 0.9561 on Tor and 0.9552 on VPN). This confirms that forcing logical rigorousness inherently enhances the overall multimodal reasoning reliability.

\subsection{Qualitative Evaluation on Traffic Reasoning}

Despite the quantitative evaluations, this section presents qualitative analysis of traffic reasoning on three datasets: ISCX-Tor-2016, CSTNet-TLS1.3, and USTC-TFC-2016.

\textbf{Qualitative Evaluation.} Tables~\ref{tab:tor_case_a}, \ref{tab:cstnet_case_a}, and \ref{tab:ustc_case_a} present high-quality correct classifications across three datasets: ISCX-Tor-2016, CSTNet-TLS1.3, and USTC-TFC-2016. Despite the diversity of encryption contexts, mmTraffic consistently produces forensically grounded reports that accurately characterize the underlying traffic behavior. The \texttt{CHAT} case from ISCX-Tor-2016 (sample \#6227) is particularly illustrative. Although the predicted ascii bucket diverges slightly from the ground truth, the generated description correctly identifies the traffic as instant messaging protocols tunneled through Tor onion routing, accurately attributing it to AIM, ICQ, Facebook Chat, Hangouts, and Skype Chat services. This suggests that the joint optimization between the traffic encoder and the LLM effectively compensates for trait-level uncertainty by grounding generation in a semantically aligned feature space. A similar pattern is observed in the \texttt{Outlook} case from USTC-TFC-2016 (sample\#11786). While the predicted entropy and zero-padding buckets differ from the ground truth, the generated description correctly identifies the traffic as Microsoft Outlook using HTTPS and MAPI, and accurately distinguishes it from Gmail by referencing Outlook's exclusive connection to Microsoft infrastructure. This decoupling between trait accuracy and description quality demonstrates that the auxiliary classification head enforces categorical boundaries in the latent space, anchoring the LLM's generation to the correct traffic identity even when individual byte-level features are imprecise. The \texttt{Steam} case from CSTNet-TLS1.3 (sample \#350) further reinforces this observation: under TLS 1.3 where all flows share identical encryption overhead, mmTraffic produces a platform-specific description referencing Valve Steam's game downloads and multiplayer services, going beyond what raw byte features alone could support.

\textbf{Qualitative Effect of Joint Optimization.} Tables~\ref{tab:tor_case_b}, \ref{tab:cstnet_case_b}, and \ref{tab:ustc_case_b} present cases where \textbf{mmTraffic} correctly identifies the traffic category while the \textit{Vanilla VLM} produces an erroneous classification, revealing how joint optimization resolves the semantic gap that a frozen encoder cannot bridge. Crucially, the misclassifications are not random: the predicted category shares substantial byte-level similarity with the ground truth. On ISCX-Tor-2016, \texttt{VIDEO} is confused with \texttt{BROWSING} because both exhibit similar Tor-tunneled TCP flows and differ primarily in sustained throughput consistency rather than observable byte patterns. On CSTNet-TLS1.3, \texttt{Adobe} is misidentified as \texttt{baidu.com} because TLS 1.3 eliminates all certificate metadata, leaving both flows with nearly identical TLS record headers and entropy profiles that are indistinguishable at the byte level with a frozen encoder. On USTC-TFC-2016, \texttt{Geodo} is confused with \texttt{Htbot} because both are HTTP-based botnets relying on similar C\&C communication patterns. This systematic pattern demonstrates that a frozen encoder fails to project fine-grained categorical boundaries into the LLM's lexical space. By contrast, mmTraffic allows gradient feedback from the auxiliary classification head to actively reshape the encoder's feature space, forcing it to learn language-aligned representations that carry explicit categorical semantics. The downstream effect is consistent: without joint optimization, the generated reports describe the wrong traffic identity, producing forensically plausible but factually incorrect outputs. 

\begin{table}[t]
\centering
\caption{ISCX-Tor-2016 for sample \#6622: mmTraffic fails to classify. 
\textcolor{red!70!black}{Red} means wrong interpretations in key semantics.}
\label{tab:tor_case_c}
\newcolumntype{G}{>{\columncolor{green!6}}p{5.5cm}}
\newcolumntype{H}{>{\columncolor{gray!6}}p{5.5cm}}
\resizebox{\linewidth}{!}{%
\begin{tabular}{l G H}
\toprule
\textbf{Field} & \textbf{mmTraffic (Ours)} & \textbf{Ground Truth} \\
\midrule
\rowcolor{gray!15}
\textbf{Class}
  & \textcolor{red!70!black}{\textbf{BROWSING}} \XSolidBrush 
  & FILE-TRANSFER \\
\midrule
\rowcolor{blue!5}
\textbf{Traits}
  & has\_tls: \texttt{false} \newline
    has\_http: \texttt{false} \newline
    ascii: \textcolor{red!70!black}{\textbf{\texttt{low}}} \newline
    entropy: \textcolor{red!70!black}{\textbf{\texttt{mid}}} \newline
    zero\_pad: \texttt{mid}
  & has\_tls: \texttt{false} \newline
    has\_http: \texttt{false} \newline
    ascii: \texttt{mid} \newline
    entropy: \texttt{low} \newline
    zero\_pad: \texttt{mid} \\
\midrule
\rowcolor{blue!5}
\textbf{Evidence}
  & (1) TCP dominant (100\%), consistent with profile. \newline
    (2) \textcolor{red!70!black}{\textbf{Unlike direct HTTPS browsing,
    Tor browsing shows only Tor relay IPs}}. \newline
    (3) \textcolor{red!70!black}{\textbf{More bursty than Tor AUDIO
    or VOIP}}. \ldots
  & (1) TCP dominant (100\%), consistent with FILE-TRANSFER
    profile. \newline
    (2) Unlike Tor P2P (bidirectional), FILE-TRANSFER is
    predominantly unidirectional. \newline
    (3) FILE-TRANSFER shows sustained throughput vs
    Tor BROWSING (bursty with idle gaps). \ldots \\
\midrule
\rowcolor{yellow!8}
\textbf{Description}
  & \textbf{\textcolor{red!70!black}{HTTPS/SSL web browsing
    tunneled through Tor onion routing}}, including both
    direct Tor Browser usage and gateway-proxied SSL
    browsing. Dominant protocol: TCP. \ldots
  & Mixed entropy and ASCII characteristics.
    \textbf{\textcolor{gray!60!black}{File transfer protocols
    (FTP, SFTP, Skype file transfer)}} tunneled through Tor
    onion routing. Dominant protocol: TCP (100\%). \\
\midrule
\rowcolor{gray!5}
\textbf{Notes}
  & Tor browsing provides strong anonymity; exit node traffic
    is unencrypted unless HTTPS is used end-to-end.
  & File transfer over Tor hides source and destination;
    commonly used to transfer sensitive documents. \\
\bottomrule
\end{tabular}%
}
\end{table}

\begin{table}[t]
\centering
\caption{CSTNet-TLS1.3 for sample \#6404: mmTraffic fails to classify. 
\textcolor{red!70!black}{Red} means wrong interpretations in key semantics.}
\label{tab:cstnet_case_c}
\newcolumntype{G}{>{\columncolor{green!6}}p{5.5cm}}
\newcolumntype{H}{>{\columncolor{gray!6}}p{5.5cm}}
\resizebox{\linewidth}{!}{%
\begin{tabular}{l G H}
\toprule
\textbf{Field} & \textbf{mmTraffic (Ours)} & \textbf{Ground Truth} \\
\midrule
\rowcolor{gray!15}
\textbf{Class}
  & \textcolor{red!70!black}{\textbf{arXiv}} \XSolidBrush 
  & Semantic Scholar \\
\midrule
\rowcolor{blue!5}
\textbf{Traits}
  & has\_tls: \texttt{true} \newline
    has\_http: \texttt{false} \newline
    ascii: \textcolor{red!70!black}{\textbf{\texttt{low}}} \newline
    entropy: \texttt{high} \newline
    zero\_pad: \texttt{mid}
  & has\_tls: \texttt{true} \newline
    has\_http: \texttt{false} \newline
    ascii: \texttt{mid} \newline
    entropy: \texttt{high} \newline
    zero\_pad: \texttt{mid} \\
\midrule
\rowcolor{blue!5}
\textbf{Evidence}
  & (1) TLS record header detected. \newline
    (2) Low ASCII with high entropy: encrypted payload. \newline
    (3) TCP dominant (96.8\%), \textcolor{red!70!black}{\textbf{consistent
    with arXiv profile}}. \newline
    (4) \textcolor{red!70!black}{\textbf{Similar PDF download pattern to
    ACM and IEEE}}. \ldots
  & (1) TLS record header detected. \newline
    (2) High Shannon entropy: encrypted payload. \newline
    (3) TCP dominant (97.9\%), consistent with Semantic
    Scholar profile. \newline
    (4) More API-driven than traditional academic publishers. \ldots \\
\midrule
\rowcolor{yellow!8}
\textbf{Description}
  & TLS record headers alongside high-entropy payload confirm
    encrypted communication.
    \textbf{\textcolor{red!70!black}{arXiv open-access academic
    preprint repository}} for physics, math, and CS over
    TLS 1.3. Large avg. packet (925 bytes). \ldots
  & TLS record headers alongside high-entropy payload confirm
    encrypted communication.
    \textbf{\textcolor{gray!60!black}{Semantic Scholar AI-powered
    academic paper search and knowledge graph}} over TLS 1.3.
    Large avg. packet (964 bytes). \ldots \\
\midrule
\rowcolor{gray!5}
\textbf{Notes}
  & Open academic repository; monitor for automated bulk
    downloading indicating research scraping bots.
  & Allen Institute AI-operated academic search; open access
    focus means fewer paywall authentication flows. \\
\bottomrule
\end{tabular}%
}
\end{table}

\begin{table}[t]
\centering
\caption{USTC-TFC-2016 for sample \#4090: mmTraffic fails to classify. 
\textcolor{red!70!black}{Red} means wrong interpretations in key semantics.}
\label{tab:ustc_case_c}
\newcolumntype{G}{>{\columncolor{green!6}}p{5.5cm}}
\newcolumntype{H}{>{\columncolor{gray!6}}p{5.5cm}}
\resizebox{\linewidth}{!}{%
\begin{tabular}{l G H}
\toprule
\textbf{Field} & \textbf{mmTraffic (Ours)} & \textbf{Ground Truth} \\
\midrule
\rowcolor{gray!15}
\textbf{Class}
  & \textcolor{red!70!black}{\textbf{Geodo}} \XSolidBrush 
  & Htbot \\
\midrule
\rowcolor{blue!5}
\textbf{Traits}
  & has\_tls: \texttt{false} \newline
    has\_http: \texttt{false} \newline
    ascii: \texttt{high} \newline
    entropy: \texttt{low} \newline
    zero\_pad: \textcolor{red!70!black}{\textbf{\texttt{mid}}}
  & has\_tls: \texttt{false} \newline
    has\_http: \texttt{false} \newline
    ascii: \texttt{high} \newline
    entropy: \texttt{low} \newline
    zero\_pad: \texttt{mid} \\
\midrule
\rowcolor{blue!5}
\textbf{Evidence}
  & (1) High ASCII with low entropy: repetitive plaintext
    content. \newline
    (2) DNS dominant (100\%), consistent with profile. \newline
    (3) Unlike Zeus (banking only), \textcolor{red!70!black}{\textbf{Geodo
    also performs spam relay}}. \ldots
  & (1) High ASCII with low entropy: repetitive plaintext
    content. \newline
    (2) DNS dominant (100\%), consistent with Htbot profile. \newline
    (3) Unlike Miuref (click fraud), Htbot focuses on C\&C
    command execution. \\
\midrule
\rowcolor{yellow!8}
\textbf{Description}
  & Substantial readable ASCII content present.
    \textbf{\textcolor{red!70!black}{Geodo/Emotet banking trojan
    and email spam botnet}}. Dominant protocol: DNS (89.1\%).
    Uses HTTP-based C\&C more frequently than Zeus. 
  & Substantial readable ASCII content present.
    \textbf{\textcolor{gray!60!black}{HTTP-based botnet using
    web proxies for C\&C (Htbot)}}. Dominant protocol:
    DNS (100\%). \\
\midrule
\rowcolor{gray!5}
\textbf{Notes}
  & High-risk banking trojan; block known Geodo C\&C IPs;
    inspect SMTP traffic for spam relay.
  & Monitor for HTTP requests with unusual headers; correlate
    with known Htbot infrastructure. \\
\bottomrule
\end{tabular}%
}
\end{table}

\subsection{Limitation Analysis}
\textbf{Failure Cases.} Tables~\ref{tab:tor_case_c}, \ref{tab:cstnet_case_c}, and \ref{tab:ustc_case_c} present failure cases where mmTraffic produces an incorrect classification. On ISCX-Tor-2016, \texttt{FILE-TRANSFER} is misclassified as \texttt{BROWSING} because both categories produce similar TCP flows under Tor multi-hop encryption, without an observable protocol marker to distinguish sustained file transfer from bursty web browsing. On CSTNet-TLS1.3, \texttt{Semantic Scholar} is misclassified as \texttt{arXiv} because both are open-access academic platforms that share nearly identical TLS 1.3 byte signatures, making them fundamentally indistinguishable at the byte level without application-layer metadata. On USTC-TFC-2016, \texttt{Htbot} is misclassified as \texttt{Geodo} because both are DNS-based botnets with nearly identical ASCII ratios, entropy profiles, and protocol distributions. In all three cases, the misclassification originates at the perceptual stage: even with end-to-end joint optimization, the encoder fails to establish sufficiently discriminative boundaries between classes that share highly similar byte-level signatures. The erroneous categorical prediction then propagates into the cognitive module, producing reports that are internally consistent with the predicted class rather than the ground truth. Despite the prediction errors, this transparency remains an operational advantage: the cognitive module faithfully reflects the perceptual judgment, making the error visible and traceable in the generated report rather than silently absorbed into an opaque label. For a network analyst, the evidence chain in the generated report can be independently verified against the raw traffic, and discrepancies between the reported behavior and observed network activity serve as a natural signal that the classification may warrant further investigation. In security-sensitive applications such as malware triage, encrypted traffic auditing, and network incident response, both high-fidelity generation under correct perception and transparent failure under incorrect perception represent a practical step toward interpretable traffic analysis.

\textbf{Paradigm, Benchmark and Evaluation.} While mmTraffic demonstrates strong performance across diverse encrypted traffic benchmarks, several aspects of the current design require further investigation. The tight coupling between the perceptual and cognitive layers 
also means that the traffic reasoning quality is inherently linked to the reliability of the traffic encoder. Exploring mechanisms that allow the cognitive layer to express uncertainty or partially recover from perceptual errors represents a promising direction for future work. Additionally, the current benchmark construction pipeline relies on \texttt{Claude Opus} to generate reference reports from structured traffic features. Although this pipeline can produce high-quality annotations, its scalability to open-world applications with new datasets or emerging traffic categories may be limited. Finally, a more comprehensive traffic reasoning evaluation protocol is still a worthwhile avenue. 

\section{Conclusion and Future Work}

This paper addresses multi-modal traffic reasoning for the first time, successfully bridging the gap between high-precision encrypted traffic classification and human-readable forensic report generation. By developing the foundational Byte-Grounded Traffic Description benchmark, i.e., \textbf{BGTD} and proposing a jointly optimized multi-modal traffic reasoning architecture with large language model (LLM), i.e., \textbf{mmTraffic}, we transform encrypted traffic analysis from a black-box classification paradigm to an auditable generative paradigm towards explainable traffic interpretations. Unlike previous decoupled pipelines that freeze the traffic encoder, mmTraffic actively unfreezes the encoder and trains it synergistically with the LLM. 
Extensive evaluations across six diverse benchmarks demonstrate that mmTraffic achieves high-fidelity, evidence-grounded traffic report generation, while maintaining highly competitive classification accuracy, confirming its success in resolving the semantic gap between physical network bytes and human-understandable concepts.

In future work, one key direction is to optimize inference latency to support real-time flow analytics and introduce uncertainty quantification to allow the cognitive layer to explicitly handle low-confidence perceptual predictions in adversarial scenarios. Furthermore, a more scalable automated annotation process to efficiently extend the framework to emerging cryptographic protocols and open-world traffic categories remains a worthwhile direction.

\appendices


%
%

\ifCLASSOPTIONcaptionsoff
  \newpage
\fi



%

\bibliographystyle{IEEEtranS}
\bibliography{sn-bibliography}
\end{document}